%% file: spanning.tex
\documentclass{iopart}

\expandafter\let\csname equation*\endcsname\relax
\expandafter\let\csname endequation*\endcsname\relax
\usepackage{amsmath,amssymb,amsthm,amsfonts}
\usepackage{marvosym}
\usepackage{wasysym}

\usepackage[english]{babel}

\usepackage{graphicx,psfrag}
\usepackage{color}
\usepackage{multirow}
\usepackage{tabu}
\usepackage{xcolor}
\usepackage[autolanguage]{numprint}
\usepackage{hyperref}
\usepackage{url}
\usepackage{diagbox}
\usepackage{algorithm}
\usepackage{algpseudocode}
\usepackage{framed}
\usepackage{pgfplots}

\newcommand{\seqnum}[1]{\href{http://oeis.org/#1}{\underline{#1}}}

\newcommand{\pstar}{\ensuremath{p_{\star}}}
\newcommand{\pmed}{\ensuremath{p_{\mathrm{med}}}}
\newcommand{\ppol}{\ensuremath{p_{\mathrm{pol}}}}
\newcommand{\pcell}{\ensuremath{p_{\mathrm{cell}}}}
\newcommand{\nmax}{\ensuremath{n_{\mathrm{max}}}}

\newcommand\Vtextvisiblespace[1][.3em]{%
  \mbox{\kern.06em\vrule height.5ex}%
  \vbox{\hrule width#1}%
  \hbox{\vrule height.5ex}}

\newcommand{\leer}{\Vtextvisiblespace[0.6em]}
\newcommand{\topc}{\ensuremath{\vert\vert}}
\newcommand{\leftc}{\mbox{((}}
\newcommand{\rightc}{\mbox{))}}
\newcommand{\middlec}{\mbox{)(}}
\newcommand{\singlec}{\mbox{()}}

\newcommand{\bothfull}{\includegraphics[width=1em]{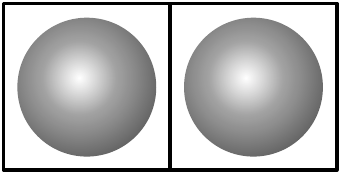}}
\newcommand{\leftfull}{\includegraphics[width=1em]{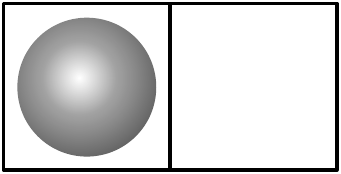}}
\newcommand{\rightfull}{\includegraphics[width=1em]{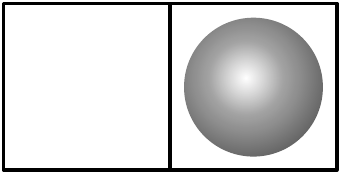}}

\begin{document}

\npthousandthpartsep{\,}

\title{Exact Percolation Probability on the Square Lattice}

\author{Stephan Mertens$^{1,2}$}

\address{$^1$Institut f\"ur Physik, Universit\"at
  Magdeburg, Universit\"atsplatz~2, 39016~Magdeburg, Germany}
\address{$^2$Santa Fe Institute, 1399 Hyde Park Rd., Santa Fe, NM 87501, USA}

\ead{mertens@ovgu.de}

\begin{abstract}
We present an algorithm to compute the exact probability
$R_{n}(p)$ for a site percolation cluster to span an $n\times n$
square lattice at occupancy $p$.  The algorithm has time and space
complexity $O(\lambda^n)$ with
$\lambda \approx 2.6$. It allows us to compute $R_{n}(p)$ up to
$n=24$. We use the data to compute estimates for the percolation
threshold $p_c$ that are several orders of magnitude more precise than estimates
based on Monte-Carlo simulations.
\end{abstract}


\input{intro}
\input{generating-function}

\input{algorithm}

\input{results}

\input{conclusions}

\appendix
\input{nsig}

\bibliographystyle{iopart-num}
\bibliography{percolation,animals,statmech,mertens,cs,math}

\end{document}

%% file: intro.tex
\section{Introduction}
\label{sec:introduction}

Most of the work in percolation is intimately connected with computer
simulations. Over the years, various algorithms have been developped
\cite{ziff:encyclopedia}. Think of the Hoshen-Kopelman algorithm to
identify clusters \cite{hoshen:kopelman:76}, the cluster growth
algorithms \cite{leath:76b,alexandrowicz:80} or the cluster hull
generating algorithm \cite{grassberger:92,ziff:92,ziff:etal:84} to
name but a few.  Another example is the the invasion percolation
algorithm, that works well independently of the spatial dimension
\cite{mertens:moore:18b} and even on hyperbolic lattices with their
``infinite''\ dimension \cite{mertens:moore:18a}.

The microcanonical union-find algorithm by Newman
and Ziff \cite{newman:ziff:00,newman:ziff:01} was a major improvement
for Monte-Carolo simulations. It is not only exceedingly efficient, 
but also simple and beautiful.

In this contribution we will introduce an algorithm for the exact
enumeration of percolating configurations in $2d$ lattices. We will
discuss this algorithm for site percolation on the
square lattice, but the method can easily be adapted to other $2d$
lattices.

Our motivation to devise such an algorithm and to run it is
twofold. First, because we think that it is always good to have exact
solutions, even for small systems. And second, because one can
use the enumeration results to compute estimates for the percolation
threshold $p_c$ that are more accurate than anything achievable with
Monte Carlo simulations.

The latter is a recent trend in percolation. Exact
estimators for small systems combined with careful extrapolations
have lead to estimates for $p_c$ with unprecedented accuracy for many
different lattices, see \cite{scullard:jacobsen:20} and references
therein.

\begin{table}
  \centering
  \begin{tabular}{lll}
    0.569(13) & MC $78\times 78$ & 1963 \cite{dean:63} \\
    0.590(10) & Series & 1964 \cite{sykes:essam:64a}\\
    0.591(1) & MC $250\times 250$ & 1967 \cite{dean:bird:67} \\
    0.593(1) & MC $256\times 256$ & 1975 \cite{levinshtein:etal:75}\\
    \numprint{0.59274}(10) & TM $10\times\infty$ & 1985
                                                   \cite{derrida:stauffer:85}\\
    \numprint{0.592745}(2) & MC $2048\times 2048$ & 1986 \cite{ziff:sapoval:86}\\
    \numprint{0.59274621}(13) & MC $128\times 128$ & 2000 \cite{newman:ziff:00}\\
    \numprint{0.59274621}(33) & MC $1594\times 1594$ & 2003 \cite{deoliveira:etal:03}\\
    \numprint{0.59274603}(9) & MC $2048\times 2048$ & 2007 \cite{lee:07}\\
    \numprint{0.59274598}(4) & MC $2048\times 2048$ & 2008 \cite{lee:08} \\
    \numprint{0.59274605}(3) & TM $16\times\infty$ & 2008 \cite{feng:deng:bloete:08}\\
    \numprint{0.59274605095}(15) & TM $22\times 22$ & 2013 \cite{yang:zhou:li:13} \\
   \numprint{0.59274605079210}(2) & TM $22\times\infty$ & 2015 \cite{jacobsen:15}
  \end{tabular}
  \caption{Some published estimates of the percolation threshold for
    site percolation on the square lattice.  TM refers to transfer
    matrix algorithms, MC to Monte-Carlo methods. See
    \cite{ziff:sapoval:86} and \cite{lee:08} for more extensive
    lists. Or consult the always up to date \textsc{WikipediA} entry on percolation thresholds,
    maintained by Bob Ziff.  
   }
  \label{tab:values}
\end{table}

The situation for site percolation is depicted in
Table~\ref{tab:values}. TM refers to transfer matrix methods as a
label for enumerative, exact solutions of small systems combined with
careful extrapolations. MC stands for Monte-Carlo results. As you can
see, in 2008 TM took over MC.  We will also use TM and add two more
values to this Table.

The paper is organized in three sections plus conclusions and
appendix.  The first section defines the task and introduces the
generating function. The second section comprises a detailed
description of the algorithm, followed by a brief
discussion of the results.  In the third section we describe the
computation of high precision values for the percolation threshold
$p_c$. Sections two and three can be read independently.
The appendix explains some of the combinatorics behind the algorithm.

%% file: generating-function.tex
\section{The Number of Percolating Configurations}
\label{sec:generating-function}

Our focus is on site
percolation on the $n \times m$ square lattice, where percolation
refers to spanning the $m$ direction. In particular we want to
compute the number $A_{n,m}(k)$ of spanning configurations with $k$
sites occupied. These numbers are conveniently organized in terms
of the generating function
\begin{equation}
  \label{eq:def-F}
  F_{n,m}(z) \equiv \sum_{k=0}^{nm} A_{n,m}(k)\ z^k\,.
\end{equation}
The spanning probability at
occupancy $p$ is given by
\begin{equation}
  \label{eq:R-F}
  R_{n, m}(p) = (1-p)^{nm}\, F_{n,m}\left(\frac{p} {1-p}\right)\,.
\end{equation}

The numbers $A_{n,m}(k)$ can be computed by enumerating all $2^{nm}$
configurations. Enumerating only the hulls of the spanning clusters
\cite{ziff:92} reduces this complexity to $O(\numprint{1,7}^{nm})$,
but even with this speedup, the maximum feasible system size is
$8\times 8$ \cite{klamser:masterthesis}.

Obviously $A_{n,m}(k)=0$ for $k < m$.  For $k=m$, the only spanning
cluster is a straight line of length $m$,
\begin{subequations}
  \label{eq:low-density}
\begin{equation}
  \label{eq:k=m}
  A_{n,m}(m) = n\,.
\end{equation}
For $k=m+1$ the spanning cluster is either a straight line with one
extra site which does not contribute to spanning, or it is a line
with a kink:
\begin{equation}
  \label{eq:k=m+1}
  A_{n,m}(m+1) = n\, {nm-m \choose 1} + 2(n-1)(m-2) \qquad (m\geq 2)\,.
\end{equation}
For $k=m+2$ a spanning configuration is either a straight line with two
free sites, a line with one kink and one free site or a line with two
small or one large kink:
\begin{align}
  \label{eq:k=m+2}
    A_{n,m}(m+2) = & n\, {nm-m \choose 2} +2 (n-2) \left(\binom{m-2}{2}+(m-2)\right) \nonumber\\
                    & + 2 (n-1) \big((n m-m-3)+(m-3) (n m-m-2)\big)\\
  & +2 (n-1) \binom{m-3}{2} \qquad\qquad (m\geq 3)\,.\nonumber
\end{align}
\end{subequations}
The formulae for $k=m+3, m+4, \ldots$ can be derived using similar
considerations, but they quickly get complicated and are not very
illuminating. And of course we are here to have the computer do this work.


In the high density regime we focus on the number of empty sites.
Obviously we need at least $n$ empty sites to prevent
spanning,
\begin{subequations}
  \label{eq:high-density}
\begin{equation}
  \label{eq:high-density-1}
  A_{n,m}(nm-k) =  \binom{nm} {mn-k}  \qquad (k < n)
\end{equation}
$n$ empty sites can block spanning only if they form a path that a
king can take from the left column of an $n\times m$ chess board to
the right column in $n-1$ moves. Let  $\mathcal{I}_m(n)$ denote the
number of such paths. Then
\begin{equation}
  \label{eq:high-density-2}
   A_{n,m}(nm-n) = \mathcal{I}_m(n)\,.
 \end{equation}
\end{subequations}
This number of royal paths would be $m 3^{n-1}$ if the king were
allowed to step on rows above and below the board.  If he cannot leave
the board, the number of paths is given by a more  intricate formula
\cite{yaqubi:etal:19a}. On a quadratic board, however, the number of
paths can be expressed by the Motzkin numbers $M_{k}$ \eqref{eq:free-signatures}:
\begin{equation}
  \label{eq:kings-paths-quadratic-board}
  \mathcal{I}_n(n) = (n+2) 3^{n-2} + 2\sum_{k=0}^{n-3}(n-k-2)
  3^{n-k-3} M_k\,.
\end{equation}
Equations \eqref{eq:low-density} and\eqref{eq:high-density} can be
used as sanity check for the algorithm. Another sanity check is
provided by a surprising parity property \cite{mertens:moore:19},
\begin{subequations}
  \label{eq:parity-main}
  \begin{equation}
    \label{eq:parity}
    F_{n,m}(-1) =  \sum_{\text{even $k$}} A_{n,m}(k) - \sum_{\text{odd $k$}} A_{n,m}(k) = (-1)^{s(n,m)}
  \end{equation}
  with
  \begin{equation}
    \label{eq:parity-s}
    s(n,m) = \left\lfloor \frac{m}{2} \right\rfloor n + \left\lceil\frac{m}{2}\right\rceil \, . 
  \end{equation}
\end{subequations}

%% file: algorithm.tex
\section{Algorithm}
\label{sec:algorithm}

If you want to devise an efficient algorithm for a computational
problem, you would be well advised to consult one of the approved
design principles of the algorithms craft \cite{noc}.  In our case the
algorithmic paradigm of choice is \emph{dynamic programming}.  The
idea of dynamic programming is to identify a collection of subproblems
and tackling them one by one, smallest first, using the solutions of
the small problems to help figure out larger ones until the original
problem is solved.

\subsection{Dynamic Programming and Transfer Matrix}

In the present case, a natural set of subproblems is the computation of
$F_{n,m}(z)$ for a given occupation pattern in the $m$th row. 
For $n=2$, for example, either the left site, the right site or both sites are occupied:
\begin{displaymath}
  F_{2,m}(z) = F_{2,m}^{\leftfull}(z)+  F_{2,m}^{\rightfull}(z) + F_{2,m}^{\bothfull}\,.
\end{displaymath}
An all empty row can not occur in spanning configurations. The $m$th
row $\leftfull$ can only represent a spanning configuration if the
$(m-1)$th row is either $\leftfull$ or $\bothfull$, hence
\begin{displaymath}
  F_{2,m}^{\leftfull}(z) = z\,F_{2,m-1}^{\leftfull}(z) +
  z\,F_{2,m-1}^{\bothfull}(z)\,.
\end{displaymath}
Complementing the equations for $F_{2,m}^{\rightfull}(z)$ and
$F_{2,m}^{\bothfull}(z)$, we can write this as matrix-vector product
\begin{equation}
  \label{eq:matrix-vector}
  \begin{pmatrix}
    F_{2,m}^{\leftfull}(z) \\[1ex]
    F_{2,m}^{\rightfull}(z) \\[1ex]
     F_{2,m}^{\bothfull}(z)
   \end{pmatrix} =
   \begin{pmatrix}
     z & 0 & z \\[1ex]
     0 & z & z \\[1ex]
     z^2 & z^2 & z^2
   \end{pmatrix}
   \begin{pmatrix}
     F_{2,m-1}^{\leftfull}(z) \\[1ex]
     F_{2,m-1}^{\rightfull}(z) \\[1ex]
     F_{2,m-1}^{\bothfull}(z)
   \end{pmatrix}\,.
\end{equation}
For general values of $n$ we can write
\begin{equation}
   \label{eq:transfermatrix}
   F_{n,m}^\sigma (z) = \sum_{\sigma'} T_{\sigma,\sigma'}(z) \,F^{\sigma'}_{n,m-1}(z)\,,
\end{equation}
where $\sigma$ and $\sigma'$ denote possible configurations that can
occur in a row of width $n$. Each $\sigma$ in row $m$ represents all
compatible configurations in the $n\times m$ lattice ``above''
it. Appropriately we refer to $\sigma$ as signature of all these
configurations.

The idea of the algorithm to compute $F_{n,1}^\sigma (z)$ for all
signatures $\sigma$ and then use \eqref{eq:transfermatrix} to
iteratively compute $F_{n,2}^\sigma(z)$, $F_{n,3}^\sigma(z)$, etc.,
thereby reducing the two-dimensional problem with complexity
$O(2^{nm})$ to a sequence of $m$ one-dimenional problems of complexity
$O(\lambda^n)$ each.
 
This particular form of a dynamic programming algorithm is known as
\emph{transfer matrix method}. When transfer matrices were introduced
into statistical physics by Kramers and Wannier in 1941
\cite{kramers:wannier:41a}, they were meant as a tool for analytical
rather than numerical computations.  Their most famous application was
Onsager's solution of the two dimensional Ising model in 1944
\cite{onsager:44a}.  With the advent of computers, transfer marices
were quickly adopted for numerical computations, and already in 1980,
they were considered ``an old tool in statistical mechanics''
\cite{derrida:vannimenus:80}. Yet the idea to use transfer matrices to
compute spanning probabilities in percolation emerged only recently,
and not in statistical physics but in theoretical computer
science. For their analysis of a two-player board game, Yang, Zhou and
Li \cite{yang:zhou:li:13} developped a dynamic programming algorithm
to compute numerically exact values of the spanning probability for a
given value of the occupation probability $p$. We build on their ideas
for the algorithm presented here.

 \subsection{Signatures}

Signatures are the essential ingredients of the algorithm. How do we
code them? And what is their number? Both questions can be answered
by relating signatures to balanced strings of parentheses.

\begin{figure}
  \centering
  \includegraphics[width=0.7\columnwidth]{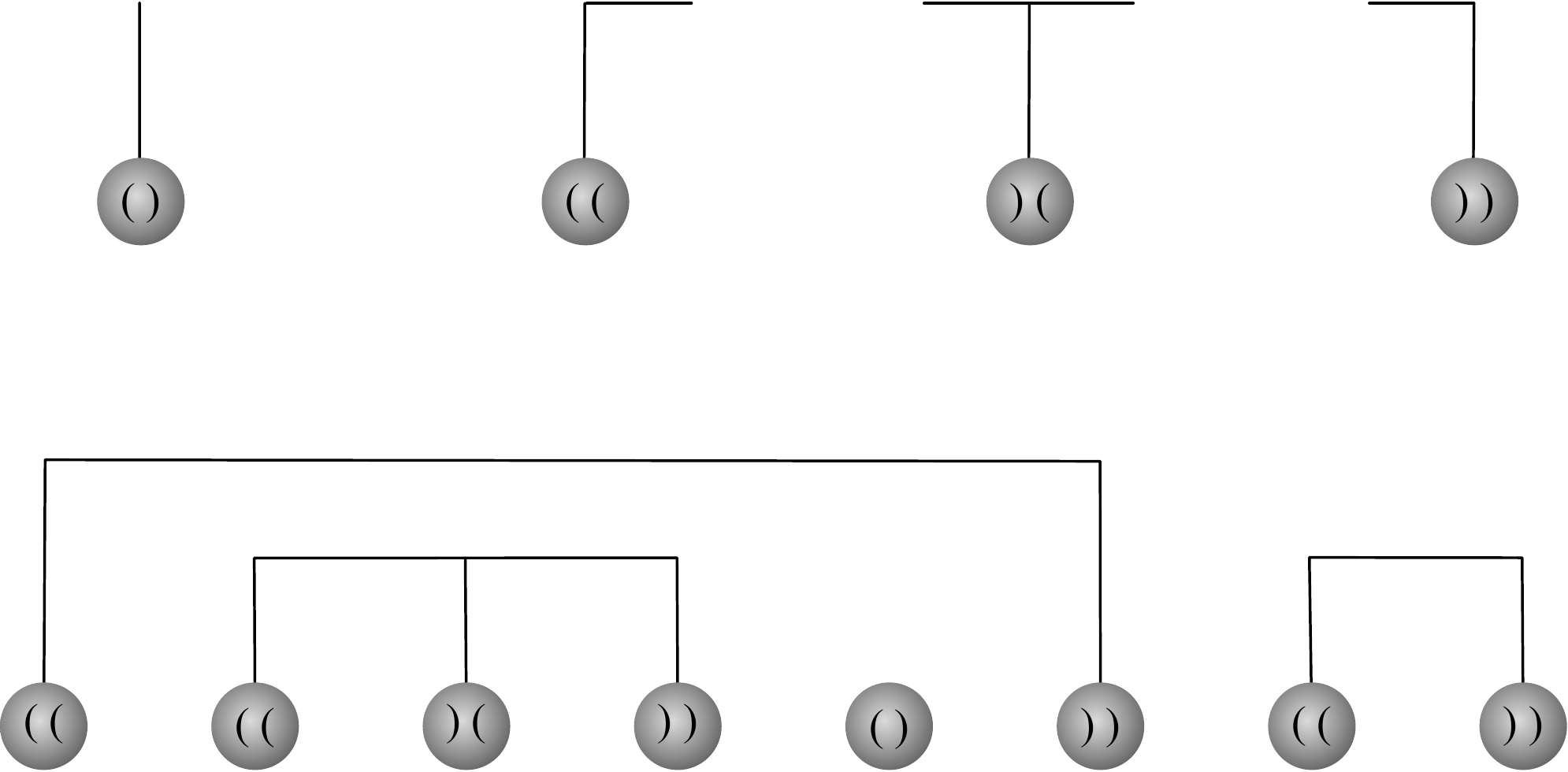}
  \caption{Coding table of cluster sites in a signature as parenthesis
    expressions (top) and example (bottom).}
  \label{fig:signature-coding}
\end{figure}

Occupied sites in a signature can belong to the same cluster. Either
because they are neighbors within the signature or because they are
connected through a path in the lattice beyond the signature. In any
case the set of signature sites that belong to the same cluster has a
unique left to right order. This allows us to encode cluster sites
using left and right parentheses (Figure~\ref{fig:signature-coding}).
We represent the leftmost site of a cluster by two left parentheses
\leftc, the rightmost site by two right parentheses \rightc\ and any
intermediate cluster site by \middlec. An occupied site that is not
connected to any other site within the signature is represented by \singlec.

The resulting string is a balanced parentheses expression (with some
white spaces from unoccupied sites), i.e., the total number of left
parentheses equals the total number right parentheses, and counting
from left to right, the number of right parentheses never exceeds the
number of left parentheses. The latter is due to the topological
constraints in two dimensions (two clusters can not cross).
Identifying all sites of a cluster in a signature means matching
parentheses in a balanced string of parentheses, a task that can
obviously be accomplished in time $O(n)$.

Since we take into account only configurations that contain at least
one spanning cluster, we need to identify clusters that are connected to the top
row.  We could follow \cite{yang:zhou:li:13} and use brackets [[, ]],
][ and [] to represent those clusters in a signature, thereby maintaining
the concept of balanced strings of parentheses. It is more efficient,
however, to use only a single symbol like \topc\ for sites that are
connected to the top. We can think of all \topc\ sites being connected
to each other by an all occupied zeroth row. Indentifying the left-
and rightmost \topc\ site in a signature can again be done in
time $O(n)$.

The total number of signatures for spanning configurations depend on
the width $n$ and the depth $m$. Some signatures can occur
only in lattices of a minimum depth. The most extreme case in this
respect are nested arcs, represented by signatures like
\begin{center}
  \leftc\, \textvisiblespace\, \leftc \, \textvisiblespace\, $\cdots$
  \, \textvisiblespace\, \rightc \, \textvisiblespace\, \rightc,
\end{center}
which can only occur if $m \geq \lceil n/2 \rceil$. If $m$ is larger
than this, however, the total number of signatures will only depend on
$n$. In the Appendix we will prove that this number is given by
\begin{equation}
  \label{eq:plain-signatures-n}
  S_n = \sum_{r=0}^{\lfloor \frac{n-1}{2}\rfloor} {{n+1} \choose
    {2r+2}}{{2r+2} \choose r}\,\frac{3}{r+3} \equiv M_{n+1,2}\,,
\end{equation}
where $M_{n+1,2}$ is the second column in the Motzkin triangle \eqref{eq:motzkin-catalan-triangle}.
The first terms of the series $S_1, S_2, \ldots$ are
\begin{displaymath}
  \label{eq:spanning-signatures-numbers}
  1, 3, 9, 25, 69, 189, 518, 1422, 3915, 10813, 29964, 83304, 232323,
  649845, \ldots 
\end{displaymath}
We also show in the Appendix, that 
$S_{n}$ grows asymptotically as
\begin{equation}
  \label{eq:Sn0-asymp}
  S_{n}\sim\frac{\sqrt{3}\,3^{n+3}}{\sqrt{4\pi
      n^3}}\left(1-\frac{159}{16 n} + \frac{36505}{512 n^2} +O(\frac{1}{n^3})\right)\,.
\end{equation}

\subsection{From one dimension to zero dimensions}

\begin{figure}
  \centering
  \includegraphics[width=0.7\linewidth]{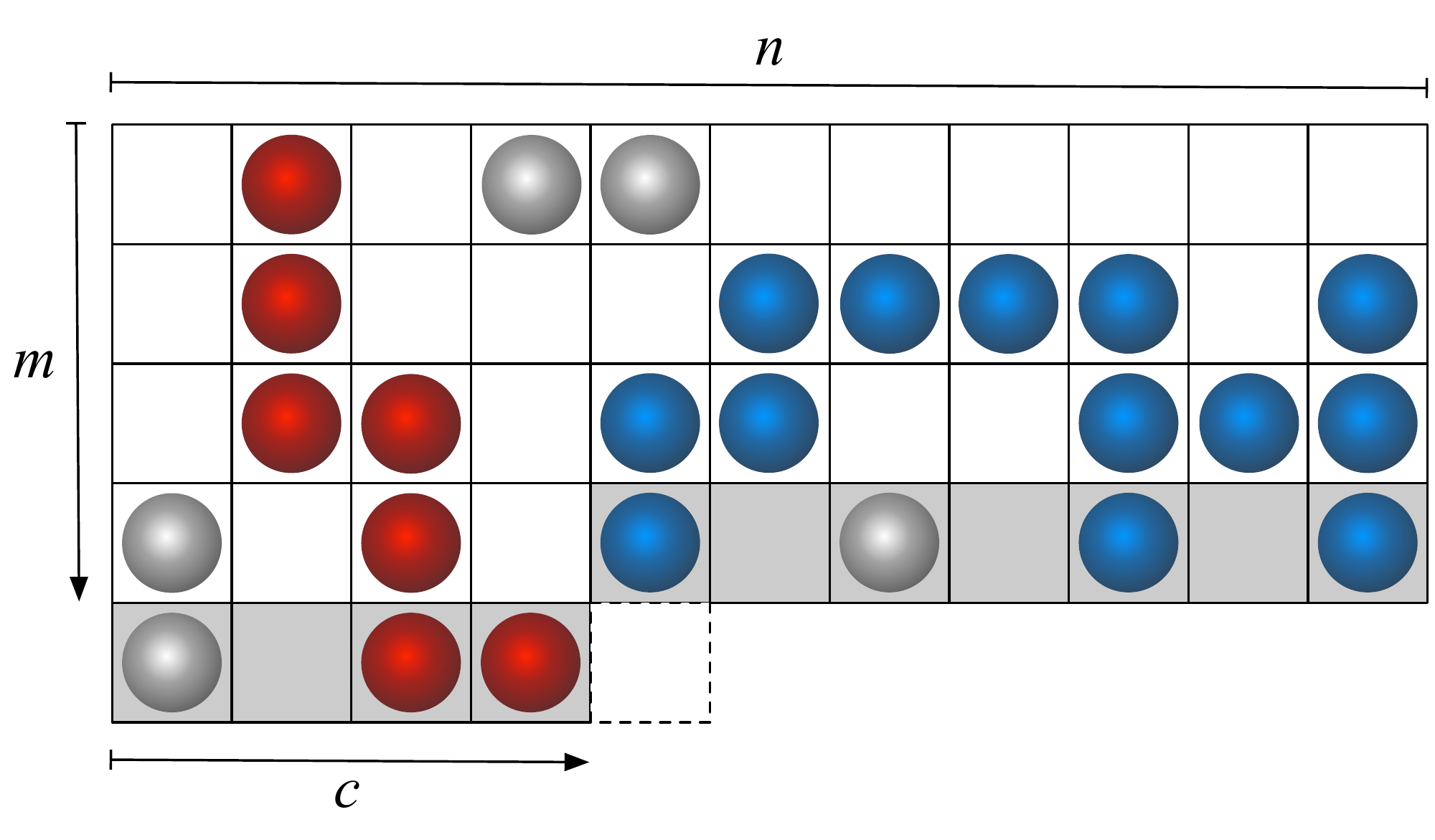}
  \caption{Increasing the system size by adding a new site in the
    $(c+1)$th column of the partially filled row. The state of the new
    site may affect the cluster structure of the whole lattice, but we
    only need to update the \emph{signature}, the cluster structure as
    it shows in the $n$ sites colored gray.}
  \label{fig:site-by-site}
\end{figure}

The central idea of the transfer matrix method
\eqref{eq:transfermatrix} is to transform a two dimensional problem into
a sequence of one dimensional problems, as a result of which the complexity
reduces from $O(2^{nm})$ to $O(m 2^n S_n)$. This idea can be applied
again. By building the new row site by site
(Figure~\ref{fig:site-by-site}), we can subdivide the one dimensional
problem into a sequence of $n$ zero dimensional problems. This reduces
the complexity from $O(m 2^n S_n)$ to $O(nm S_n)$.

\begin{figure}
  \centering
  \includegraphics[width=0.6\linewidth]{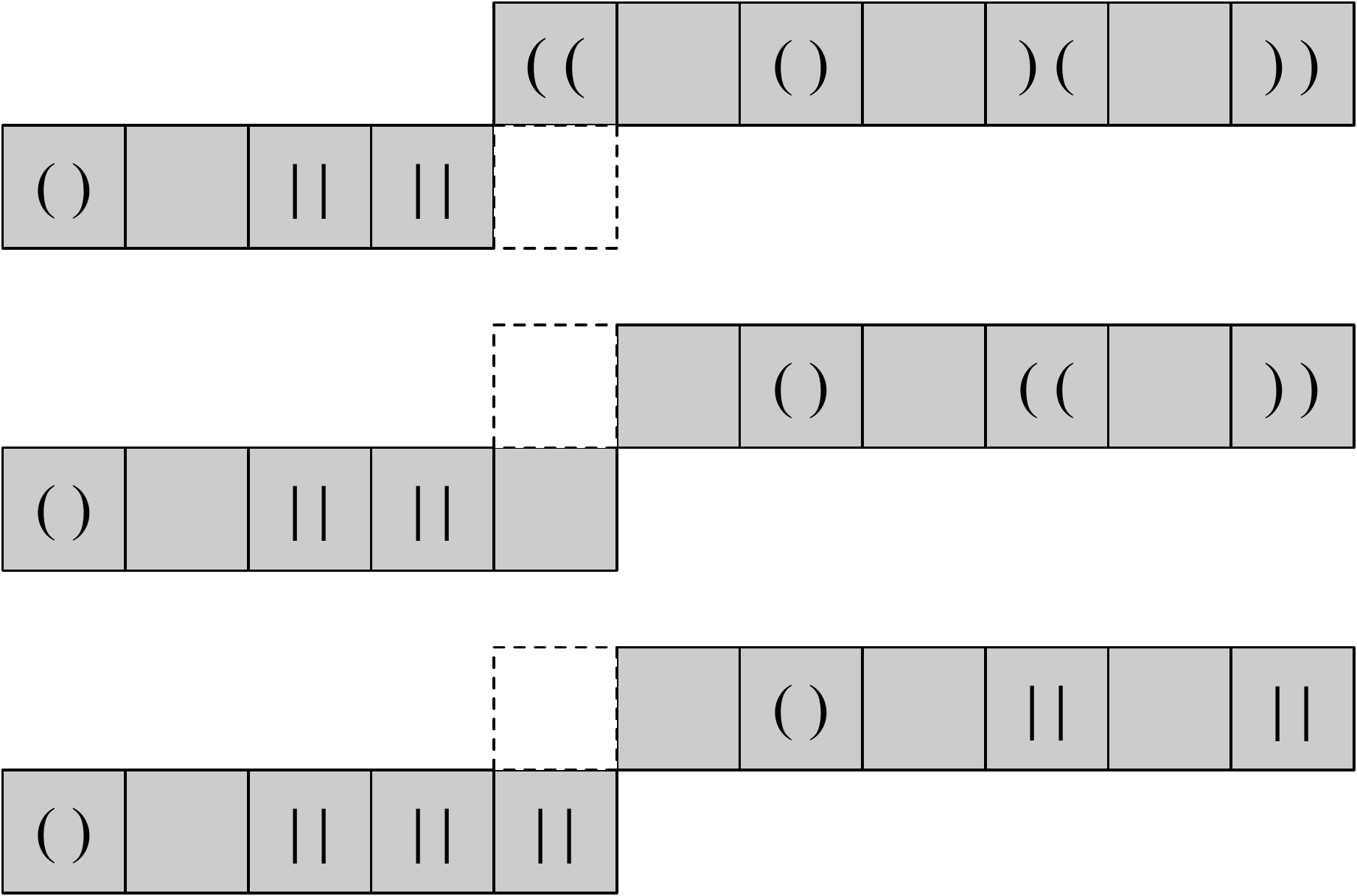}
    \caption{Signature of the cluster configuration of
    Fig~\ref{fig:site-by-site} (top), and signatures after adding a new
    site that is either empty (middle) or occupied (bottom).}
  \label{fig:signature-update}
\end{figure}

In this approach, the new row is partially filled up to column
$c$. The corresponding signatures contain a kink at column $c$. 
Each signature is mapped to two signatures, depending on whether the
added site at columns $c+1$ is empty or occupied
(Figure~\ref{fig:site-by-site}).

The number of signatures $S_{n,c}$ depends on $n$ and $c$.  We can think of a
partially filled row as a full row of width $n+1$, where an extra site
has been inserted between positions $c$ and $c+1$, this extra site
being empty in all signatures. This gets us
\begin{equation}
  \label{eq:partial-signatures-count}
  S_{n} \leq S_{n,c} \leq S_{n+1}\,,
\end{equation}
where the first inequality reflects the fact that the site at position
$c$ has only two neighbors (left and up), and therefore less
restrictions to take on ``legal''\ values. The actual value of
$S_{n,c}$ depends on $c$, see Table~\ref{tab:partial}. 

\begin{table}
  \centering
  \setlength{\tabcolsep}{3pt}
  \begin{tabular}{c|rrrrrrrr}
 \diagbox[innerrightsep=5pt]{$n$}{$c$}  & 1 & 2 & 3 & 4 & 5 & 6 & 7 & 8\\\hline
1 & \rule{0pt}{2.5ex} $1$ &  & & & & & & \\
2& 4 & 3 & & & & & & \\
3& 12 &  11 & 9 & & & & & \\
4& 33 & 31 & 31 & 25 & & & & \\
5& 91 & 85 & 87 & 85 & 69 & & &  \\
6& 249 & 233 & 237 & 237 & 233 & 189 & & \\
7& 683 & 638 & 651 & 646 & 651 & 638 & 518 & \\
8& 1877 & 1751 & 1788 & 1777 & 1778 & 1787 & 1752 & 1422
  \end{tabular}
  \caption{Values of $S_{n,c}$.}
  \label{tab:partial}
\end{table}

\subsection{Condensation}

\begin{figure}
  \centering
  \includegraphics[width=0.7\linewidth]{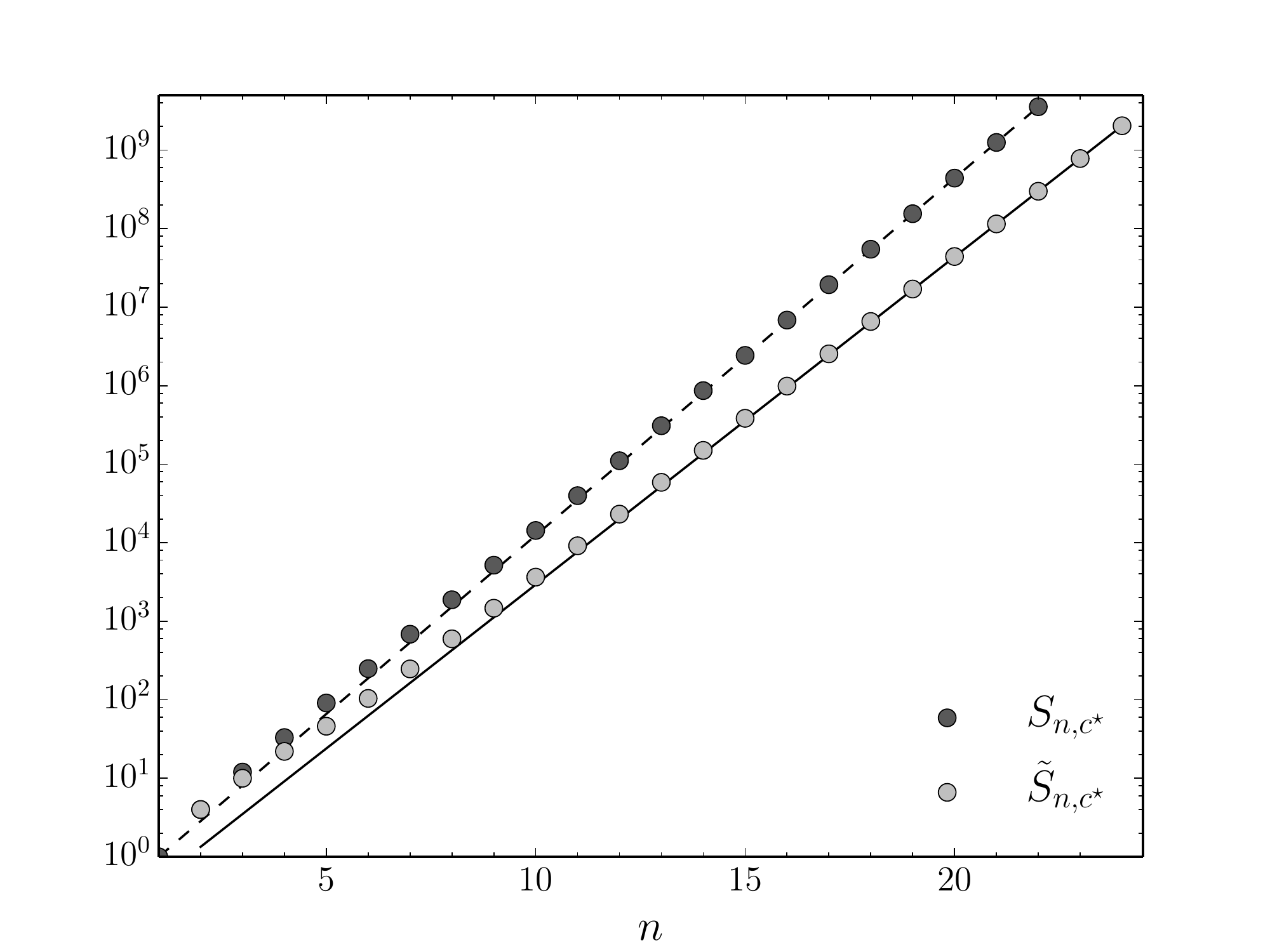}
    \caption{Number of signatures $S_{n,c^\star}$ (plain) and
      $\tilde{S}_{n,c^\star}$ (condensed). The lines are $S_{n,c^\star}\simeq
      \numprint{0.350}\cdot\numprint{2.85}^n$ and $\tilde{S}_{n,c^\star}\simeq
      \numprint{0.196}\cdot\numprint{2.61}^n$.}
  \label{fig:signature-count}
\end{figure}

The number of signatures is paramount for the complexity of
the algorithm. Let
$S_{n,c^\star} \equiv \max_c
S_{n,c}$. Figure~\ref{fig:signature-count} shows that
$S_{n,c^\star}\simeq\numprint{0.350}\cdot\numprint{2.85}^n$. In
particular, $S_{n,c^\star}$ exceeds the ``Giga''\ threshold
($10^9$) for $n > 20$. This is significant because we need to store
pairs $(\sigma, F^\sigma)$ for all signatures. For $S_{n,c^\star}$
above $10^9$, this means hundreds of GBytes of RAM. In this regime, space (memory)
rather than time will become the bottleneck of our algorithm.

It turns out, however, that we can reduce the number of
signatures. Consider a signature in row $m$ that contains the pattern
$\ldots\leer\,\singlec\,\leer\ldots$. All signatures in row $m+1$ that
are derived from it, do not depend on whether the single $\singlec$ is
actually occupied. Hence we can identify all signatures
$\sigma = \ldots\leer\,\singlec\,\leer\ldots$ with signatures
$\sigma' = \ldots\leer\,\leer\,\leer\ldots$ and store only one pair
$(\sigma', F^\sigma + F^{\sigma'})$ instead of two pairs $(\sigma,
F^\sigma)$ and $(\sigma', F^{\sigma'})$.
For the same reason, we can also identify
$\ldots\leer\;\leftc\;\rightc\;\leer\ldots$ with
$\ldots\leer\;\leer\;\leer\;\leer\ldots$
and $\ldots\leer\;\leftc\;\middlec\;\rightc\;\leer\ldots$ with
$\ldots\leer\;\leftc\;\leer\;\rightc\;\leer\ldots$
and store only one of each pair.

In addition, the symmetry of the problem allows us to identify each
signature of a full row ($c=n$) with its mirror image. 

Let $\tilde{S}_{n,c}$ denote the number of signatures that haven been
condensed by these rules, and let $\tilde{S}_{n,c^\star}=\max_c
\tilde{S}_{n,c}$.
Figure~\ref{fig:signature-count} shows that
$\tilde{S}_{n,c^\star}\simeq\numprint{0.196}\cdot\numprint{2.61}^n$.
The number of condensed signatures 
exceeds the ``Giga''\ threshold only for $n > 23$.


\subsection{Implementation}

\begin{table}
  \centering
  {\small \tabulinesep=1.2mm
  \begin{tabu}{cc|c|c|c|c|c|c|}
    \multicolumn{2}{c}{} & \multicolumn{6}{c}{upper} \\[0.5ex]
     \multirow{14}{*}{left} & \multicolumn{1}{c}{} & \multicolumn{1}{c}{\leer} & \multicolumn{1}{c}{\topc} & \multicolumn{1}{c}{\leftc} & \multicolumn{1}{c}{\rightc} & \multicolumn{1}{c}{\middlec} & \multicolumn{1}{c}{\singlec} \\\cline{3-8}
    &\multirow{2}{*}{\leer}  & \leer\;\leer & \leer\;\leer$^a$ & \leer\;\leer$^b$ & \leer\;\leer$^b$& \leer\;\leer & \leer \;\leer\\
      &       & \leer\;\singlec & \leer\;\topc & \leer\;\leftc & \leer\;\rightc& \leer\;\middlec & \leer\;\singlec\\\cline{3-8}
    &\multirow{2}{*}{\topc} & \topc\;\leer& \topc\;\leer & \topc\;\leer$^b$ & \multirow{2}{*}{{\large\Lightning}}  & \multirow{2}{*}{{\large\Lightning}} & \topc\;\leer \\
         &     & \topc\;\topc& \topc\;\topc & \topc\;\topc$^c$ &   &  & \topc\;\topc \\\cline{3-8}
    &\multirow{2}{*}{\leftc} & \leftc\;\leer & \multirow{2}{*}{{\large\Lightning}} & \leftc\;\leer$^b$ & \singlec\;\leer & \leer\;\leer & \leftc\;\leer\\
      &        & \leftc\;\middlec & & \leftc\;\middlec$^d$ & \leftc\;\rightc & \leftc\;\middlec & \leftc\;\middlec \\\cline{3-8}
   &\multirow{2}{*}{\rightc} & \rightc\;\leer & \rightc\;\leer$^a$ & \rightc\;\leer$^b$ & \rightc\;\leer$^b$ & \rightc\;\leer & \rightc\;\leer\\
     &         & \middlec\;\rightc & \topc\;\;\topc$^c$ & \middlec\;\middlec & \middlec\;\rightc$^d$ & \middlec\;\middlec$^d$ & \middlec\;\middlec\\\cline{3-8}
    &\multirow{2}{*}{\middlec} & \middlec\;\leer & \multirow{2}{*}{{\large\Lightning}} & \middlec\;\leer$^b$ & \rightc\;\leer & \middlec\;\leer & \middlec\;\leer \\
      &         & \middlec\;\middlec & & \middlec\;\middlec$^d$& \middlec\;\rightc & \middlec\;\middlec & \middlec\;\middlec\\\cline{3-8}
    & \multirow{2}{*}{\singlec} & \singlec\;\leer & \singlec\;\leer$^a$ & \singlec\;\leer$^b$ & \singlec\;\leer$^b$ & \singlec\;\leer & \singlec\;\leer$^b$ \\
     & &\leftc\;\rightc & \topc\;\topc & \leftc\;\middlec & \middlec\;\rightc & \middlec\;\middlec & \leftc\;\rightc \\\cline{3-8}
  \end{tabu}}
  \caption{Extension rules of the transfer matrix algorithm. Each cell shows the resulting configuration if the added site is empty (top) or occupied (bottom). Some update rules are \emph{non-local}: $^a$\topc\ may only be deleted, if there is at least one other \topc\  elsewhere in the signature, $^b$deleting \leftc\ or \rightc\ means shortening of the corresponding cluster, $^c$the whole cluster has to be connected to the top, and $^d$two clusters merge. Configurations with \Lightning\ must never occur.}
  \label{tab:rules}
\end{table}

The elementary step of the transfer matrix algorithm is the addition of a new
lattice site in a partially filled row
(Figure~\ref{fig:signature-update}).  This is implemented in a subroutine
$\textbf{extend}(\sigma, c)$ that takes a signature $\sigma$ and a
column $c$ as input and returns a pair $\sigma_0,\sigma_1$, where
$\sigma_0$ ($\sigma_1$) follows from $\sigma$ by adding an empty
(occupied) site in column $c$. The rules for these computations are
depicted in Table~\ref{tab:rules}. The signatures $\sigma_0$ and
$\sigma_1$ are of course condensed before they are returned.
The subroutine \textbf{extend} is then called from within a loop over
rows and columns of the lattice (Figure~\ref{fig:loop}). 

\begin{figure}
  \centering
  \fbox{\parbox{0.68\linewidth}{
  \begin{algorithmic}[0]
    \State $L_{\text{old}} := \text{first row configurations}$
    \State $L_{\text{new}} := \text{empty list}$
  \For{$r = 1,\ldots,m$} 
  \For{$c = 1,\ldots,n$}
  \While{$L_{\text{old}}$ not empty}
  \State take $(\sigma,F^\sigma)$ out of $L_{\text{old}}$
  \State $\sigma_0, \sigma_1 := \textbf{extend}(\sigma, c)$
  \If{$(\sigma_0, \cdot) \not\in L_{\text{new}}$} add $(\sigma_0, F^\sigma)$ to $L_{\text{new}}$
  \Else\ replace $(\sigma_0, F)$ in $L_{\text{new}}$ with $(\sigma_0,F+F^\sigma)$
  \EndIf
  \If{$(\sigma_1, \cdot) \not\in L_{\text{new}}$} add $(\sigma_0, z F^\sigma)$ to $L_{\text{new}}$
  \Else\ replace $(\sigma_1, F)$ in $L_{\text{new}}$ with $(\sigma_1,F+z F^\sigma)$
  \EndIf
  \EndWhile
  \State swap names $L_{\text{old}} \leftrightarrow L_{\text{new}}$
  \EndFor
  \EndFor
  \State $F_{n,m} := \sum_{(\sigma,F^\sigma)\in L_{\text{old}}} F^\sigma$
\end{algorithmic}}}
\caption{Transfer Matrix Loop.}
\label{fig:loop}
\end{figure}

The algorithm maintains two lists $L_{\text{old}}$ and
$L_{\text{new}}$ of configurations, where a configuration consists of
a signature $\sigma$ and the corresponding generating function
$F^\sigma$. $L_{\text{old}}$ is initialized with the $2^n-1$ first row
configurations, in which each occupied site is connected to the top row
by definition. $L_{\text{new}}$ is initially empty.

A configuration $(\sigma, F^\sigma)$ is removed from $L_{\text{old}}$
and $\sigma_0, \sigma_1=\textbf{extend}(\sigma,c)$ is computed.
Then the configurations $(\sigma_0, F^\sigma)$ and
$(\sigma_1, zF^\sigma)$ are inserted into $L_{\text{new}}$.
This is repeated until $L_{\text{old}}$ is empty.
At this point, $L_{\text{old}}$ and $L_{\text{new}}$ swap their names,
and the process can be iterated to add the next site to the lattice.

If the lattice has reached its final size $nm$, the total generating
function is the sum over all $F^\sigma$ for
$(\sigma, F^\sigma)\in L_{\text{old}}$.

With its six symbols \leer, \topc, \leftc, \middlec, \rightc and \singlec, a
signature can be interpreted as a senary number with $n$ digits.
For $n\leq 24$, this number always fits into a 64-bit integer.
Since
\begin{equation}
  \label{eq:senary}
  2^{64} = \numprint{3520522010102100444244424}_6\,,
\end{equation}
we can also represent signatures of size $n=25$ by 64-bit integers,
provided we choose the senary representation wisely: $\middlec = 4_6$ and
$\rightc = 5_6$.

The lists $L$ maintained by the algorithm contain up to
$\tilde{S}_{n,c^\star}$ configurations.  An efficient data structure
is mandatory. We take advantage of the fact that the signatures can be
ordered according to their senary value. This allows us to use an
ordered associative container like \texttt{set} or \texttt{map} from
the standard C++ library. Those guarantee logarithmic complexity
$O(\log \tilde{S}_{n,c^\star})$ for search, insert and delete
operations\cite{clr:algorithms}

\begin{figure}
  \centering
  \includegraphics[width=0.7\columnwidth]{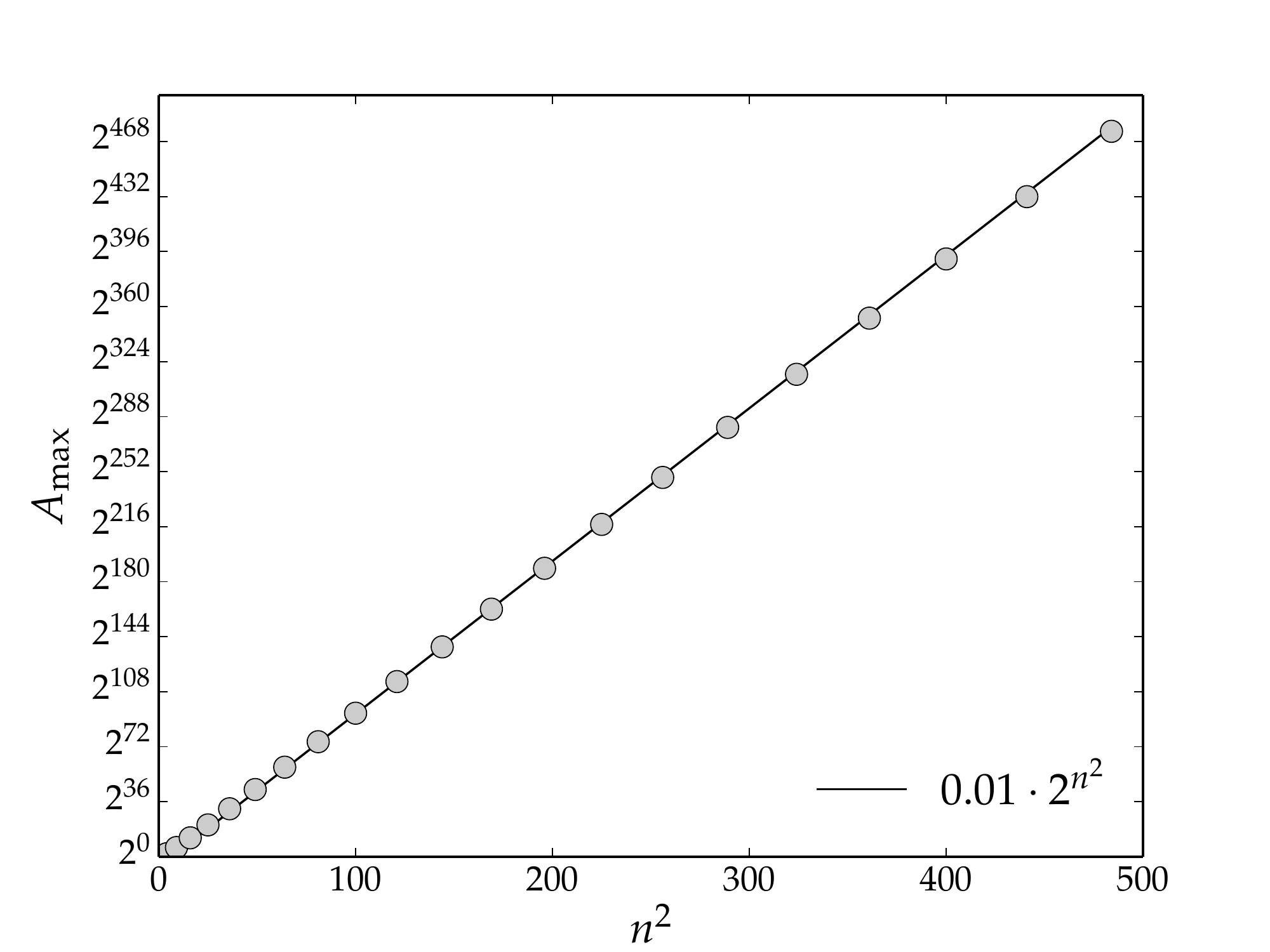}
  \caption{Maximum coefficient $A_{\mathrm{max}}(n)=\max_k A_{n,n}(k)$.}
  \label{fig:Amax}
\end{figure}

Equipped with a compact representation for the signatures and an
efficient data structure for the list of configurations, we still face the challenge to store
the large coefficients $A_{n,m}(k)$ of the generating functions
$F_{n,m}^\sigma$. These numbers quickly get too large to fit within
the standard fixed width integers with $32$, $64$ or $128$ bits
(Figure~\ref{fig:Amax}).


One could use multiple precision libraries to deal with this problem,
but it is much more efficient to stick with fixed width integers and
use modular arithmetic: with $b$-bit integers,  we 
compute $U^{(i)}_{n,m}(k) \equiv A_{n,m}(k) \bmod p_i$ for a set of prime moduli $p_i
< 2^b$ such that $\prod_{i} p_i > \max_k A_{n,m}(k)$ .
We can then use the Chinese Remainder Theorem
\cite{clr:algorithms,knuth:2} to recover $A_{n,m}(k)$ from the $U^{(i)}_{n,m}(k)$.
This way we can trade space (bit size of numbers) for time (one run for each
prime modulus). 

\subsection{Performance}

\begin{figure}
  \centering
  \includegraphics[width=0.7\linewidth]{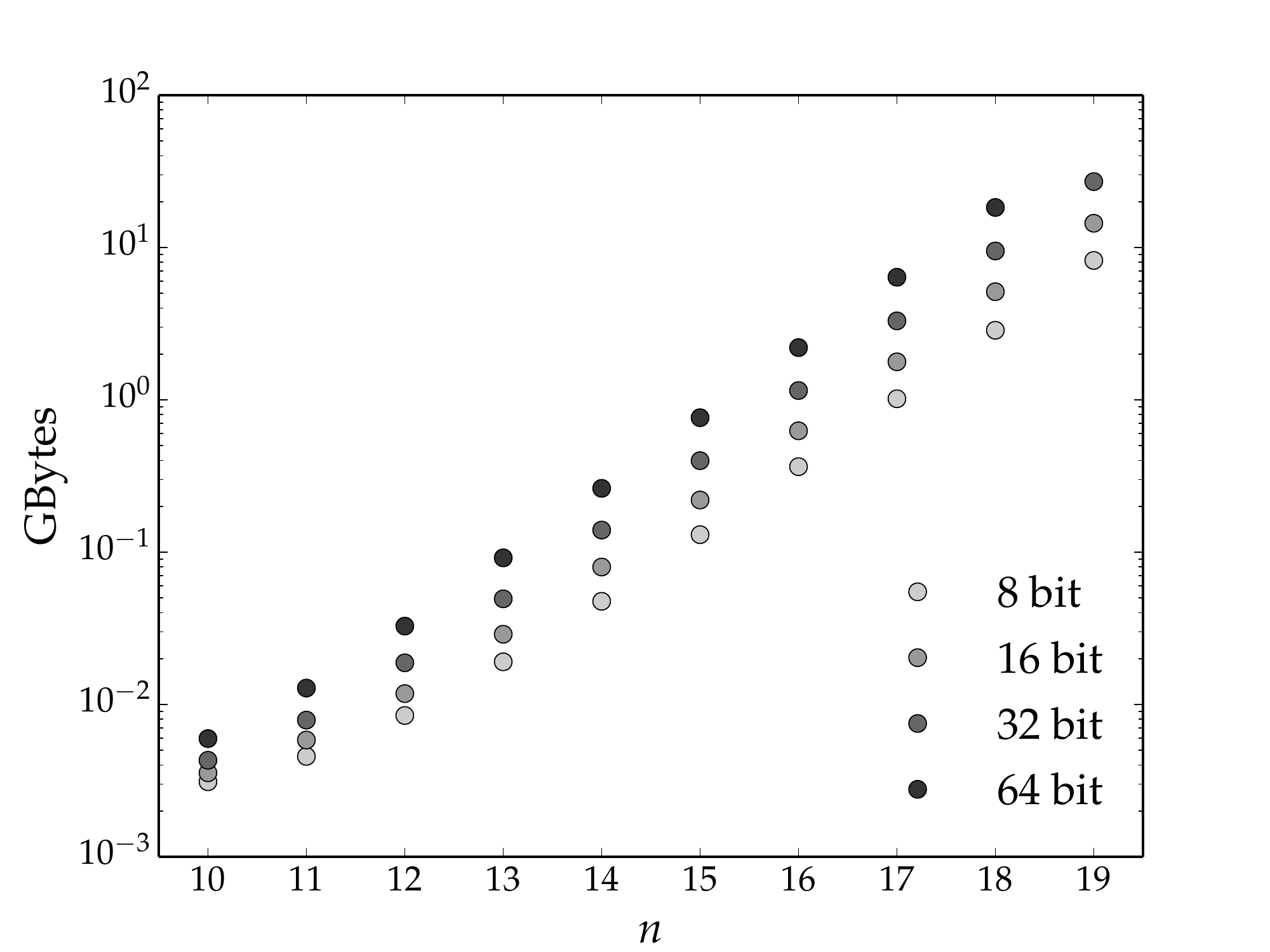}
  \caption{Memory required to compute $F_{n,n}$ with $b$-bit
    integers.}
  \label{fig:memory}
\end{figure}

Figure~\ref{fig:memory} shows the actual memory consumption to compute
$F_{n,n}$ with $b$-bit integers. Note that data structures like
red-black-tress require extra data per stored element. Unfortunately
the product of all primes below $2^8$ is not large enough to compute
the coefficients in $F_{n,n}$ via modular arithmetic and the Chinese
Remainder Theorem if $n\geq 19$. Hence 8-bit integers are essentially
useless.

Extrapolation of the data in Figure~\ref{fig:memory} shows that the
computation of $F_{n,n}$ for $n\leq 21$ cane be done with 16-bit
integers on machines with 256 GB of RAM.  Larger lattices require 384
GB ($n=22$) or 1 TB ($n=23$).

There is an alternative approach to determine $F_{n,m}$ with less
memory. Instead of enumerating the coefficients $A_{n,m}(k)$, one can
compute $F_{n,m}(z)$ for $nm$ integer arguments $z$ (again modulo a
sufficient number of primes) and then use Lagrange interpolation to
recover $F_{n,m}$.  This approach requires only a single $b$-bit
integer to be stored with each signature. The prize we pay is that now
we need $nm$ computations, each consisting of several runs for a
sufficient number of prime moduli.  This was the approach we used to
compute $F_{22,22}$ on $30\ldots 60$ machines with 256 GB each.

Yet another method suggests itself if one is mainly interested in
evaluating the spanning probability $R_{n,m}(p)$ for a given value of
$p$. Here we need to store only a single real number with each
signature, and there is no need for multiple runs with different prime
moduli. Quadruple precision floating point variables with 128 bit
(\texttt{\_\_float128} in gcc) allow us to evaluate $R_{n,m}(p)$ with
more than 30 decimals precision. We used this approach to
compute $R_{n,n}(p)$ for $n=23,24$, see Table~\ref{tab:pmedcell}. 

\begin{figure}
  \centering
  \includegraphics[width=0.7\linewidth]{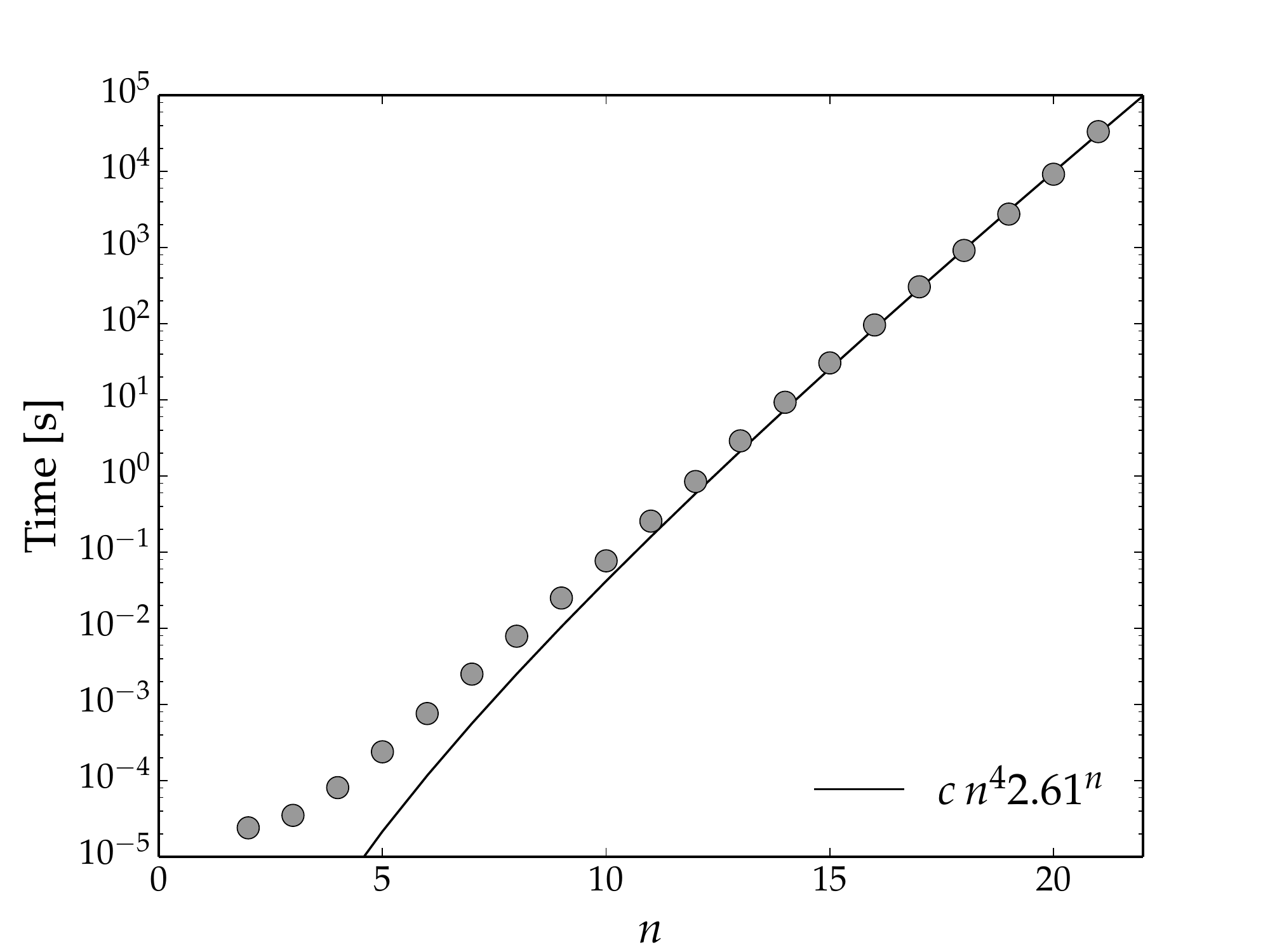}
  \caption{Time to evaluate $R_{n,n}(p)$ with 30 decimals accuracy on a
  laptop.}
  \label{fig:wallclock}
\end{figure}

Figure~\ref{fig:wallclock} shows the time 
to evaluate $R_{n,n}(p)$ for given value $p$ and with 30 decimals
accuracy, measured on a laptop. The time complexity corresponds to the expected complexity
$O(n^4 \tilde{S}_{n,c^\star})$. The $n^2$ iterations to build the
lattice, combined with the $O(n)$ complexity of the subroutine \textbf{extend}
and the $O(\log \tilde{S}_{n,c^\star}) = O(n)$ complexity of the
associative container \texttt{set} yield a scaling factor $n^4$. The
number of signatures
$\tilde{S}_{n,c^\star}\simeq\numprint{0.196}\cdot\numprint{2.61}^n$
then provide the exponential part of the time complexity.

Despite the exponential complexity, the algorithm is surprisingly fast
for small values of $n$, even on a laptop. Systems up to $12\times 12$
can be computed in less than one second, $16\times 16$ takes 90
seconds. The largest system that fits in the 16 GByte memory of the author's
laptop is $21\times 21$. A single evaluation of $R_{21,21}(p)$ on the
laptop takes about 9 hours.

\subsection{Results}

We have computed the generating functions $F_{n,m}(z)$ for $n \leq 22$
and all $m \leq n$. The data is available upon request. Here we will briefly
discuss some properties of the coefficients $A_{n,m}(k)$. For simplicity
we will confine ourselves to the quadratic case $m=n$.

$A_{nn}(k)$ has a maximum $A_{\text{max}}$ for
$k=k_{\text{max}}(n)$.  Our data reveals that  
\begin{equation}
  \label{eq:max}
  k_{\text{max}}(n) = \begin{cases}
    \frac{1}{2} n (n+1) & (n \leq 4) \\
    \frac{1}{2} n (n+1) - 1 & (4 < n \leq 22)
  \end{cases}
\end{equation}
It is tempting to conjecture that $k_{\text{max}}(n)=\frac{1}{2}n(n+1)
- O(\log n)$.

$A_{\text{max}}$ itself scales like $\simeq 2^{n^2}$
(Figure~\ref{fig:Amax}). This number determines
how many prime moduli of fixed width we need to compute $A_{n,n}(k)$
via the Chinese Remainder Theorem.

\begin{figure}
  \centering
  \includegraphics[width=0.7\columnwidth]{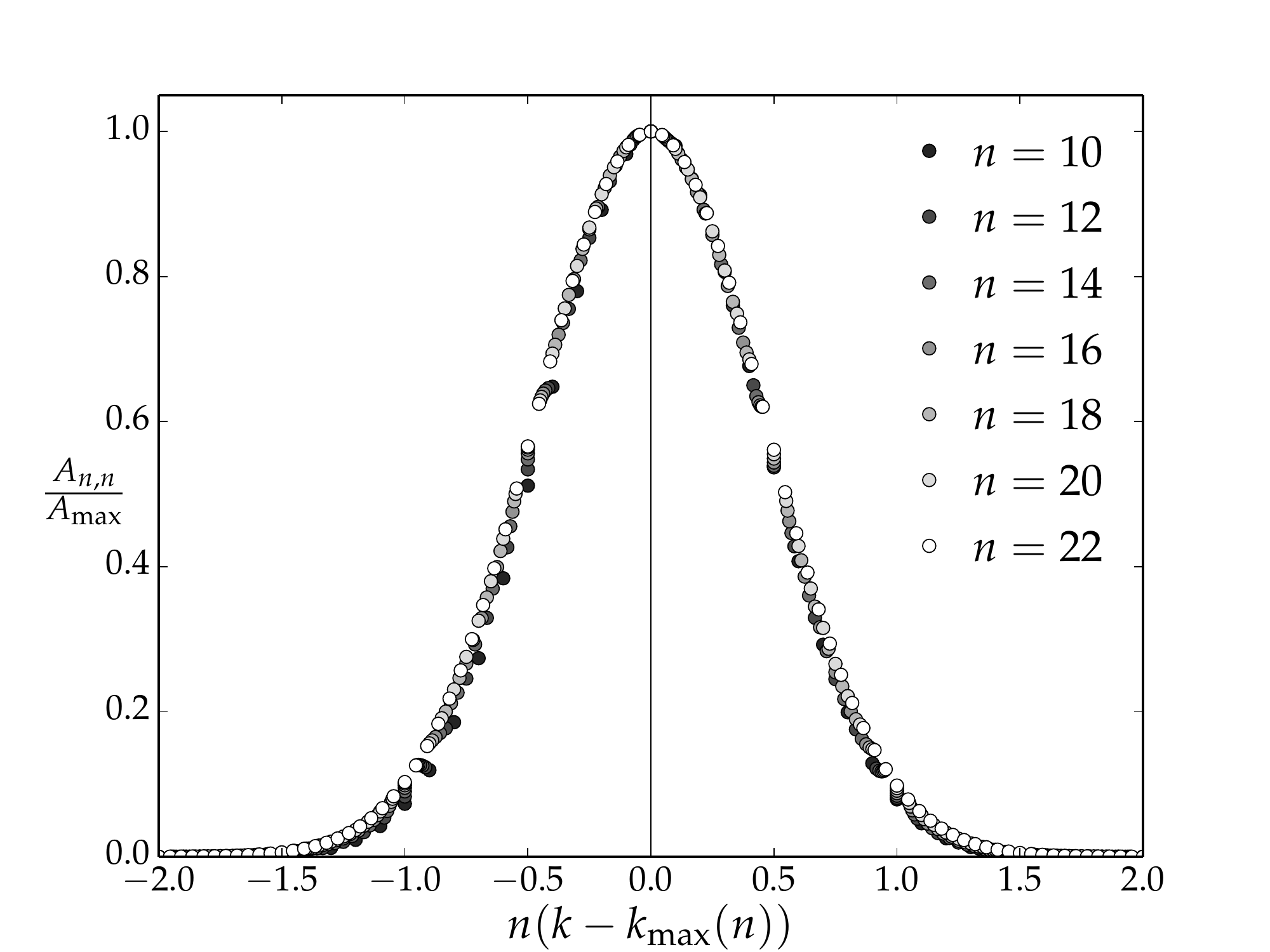}
  \caption{``Data collapse''\ in the number of spanning configurations.}
  \label{fig:scaled_A}
\end{figure}

The width of the maximum at $k_{\text{max}}$ scales like $n^{-1}$, as
can be seen in the ``data collapse''\ when plotting
$\frac{A_{n,n}(k)}{A_{\text{max}}}$
versus $n(k-k_{\text{max}})$ (Figure~\ref{fig:scaled_A}).

%% file: results.tex
\section{Critical Density}
\label{sec:results}

The generating functions $F_{n,m}(z)$ can be analyzed in various
ways. One can, for example, investigate their properties for
arguments outside the ``physical''\ interval $z \geq 0$.  This is how
the parity property \eqref{eq:parity-main} was discovered and then proven
for a large, quite general class of lattices
\cite{mertens:moore:19}.

In this paper we use the generating function to compute finite size
estimators for the critical density $p_c$ that we can then extrapolate
to the infinite lattice.

\subsection{Estimators}

The properties of various finite-size threshold estimators for 
two-dimensional percolation have been discussed by Ziff and Newman
\cite{ziff:newman:02}. Here we will focus on estimators $\pmed(n)$
and $\pcell(n)$, defined as
\begin{subequations}
  \label{eq:px_def}
  \begin{equation}
    \label{eq:pmed_def}
    R_{n,n}(\pmed) = \frac{1}{2}
  \end{equation}
  and
  \begin{equation}
    \label{eq:pcell_def}
    R_{n,n}(\pcell) = R_{n-1,n-1}(\pcell)
  \end{equation}
\end{subequations}
The values for these estimators up to $\nmax=24$ are shown in
Table~\ref{tab:pmedcell}.

\begin{table}
  \centering
  \begin{tabular}{rll}
    $n$ & $\pmed(n)$ & $\pcell(n)$ \\[1ex]
  1 & \numprint{0.500000000000000000000000000000} & \\
  2 & \numprint{0.541196100146196984399723205366} & \numprint{0.618033988749894848204586834365}\\
  3 & \numprint{0.559296316001323534755763920789} & \numprint{0.620734471730213489279430523252}\\
  4 & \numprint{0.569724133968027153228040518353} & \numprint{0.619583777217655117827765082346}\\
  5 & \numprint{0.575810073211627653605032974314} & \numprint{0.613506053696798307677288581191}\\
  6 & \numprint{0.579702757132443521419439978330} & \numprint{0.609208761667991711639989601974}\\
  7 & \numprint{0.582351295080082980073474691830} & \numprint{0.606075989265451882260569905672}\\
  8 & \numprint{0.584241466489847673860351132398} & \numprint{0.603762879768292280370335551749}\\
  9 & \numprint{0.585641556861396511416995666354} & \numprint{0.602011983338557157018606681442}\\
10 & \numprint{0.586710034053406359804690473124} & \numprint{0.600656365889941520514281630459}\\
11 & \numprint{0.587545601376707076865096376747} & \numprint{0.599585402842798543263268780420}\\
12 & \numprint{0.588212470606443263171973079741} & \numprint{0.598724257102302868949743766602}\\
13 & \numprint{0.588753953651382767097855532073} & \numprint{0.598021063882979665779438359439}\\
14 & \numprint{0.589200171193723644344640059478} & \numprint{0.597439041437080848283968950089}\\
15 & \numprint{0.589572624709616129448171107692} & \numprint{0.596951544750954799494938258995}\\
16 & \numprint{0.589887013723258987069435195722} & \numprint{0.596538896774607564321149432769}\\
17 & \numprint{0.590155029961590857817809334101} & \numprint{0.596186311743107211382955497774}\\
18 & \numprint{0.590385533260670477887261817585} & \numprint{0.595882504071658438851295137696}\\
19 & \numprint{0.590585341987611428265706601016} & \numprint{0.595618737176226408186670067141}\\
20 & \numprint{0.590759776378574624089782736423} & \numprint{0.595388160640890684818467256078}\\
21 & \numprint{0.590913039569092174937740096798} & \numprint{0.595185340192645002572869419257}\\
22 & \numprint{0.591048489639675518303675880577} & \numprint{0.595005919018756026574538302404}\\
23 & \numprint{0.591168837021863732985091670808} & \numprint{0.594846370109520325790777710845}\\
24 & \numprint{0.591276289864951685617852112360} & \numprint{0.594703812696743490456949711289}  
  \end{tabular}
 \caption{Finite size estimators for $p_c$ \eqref{eq:px_def}. All decimals are exact.}
  \label{tab:pmedcell}
\end{table}

\begin{figure}
  \centering
  \includegraphics[width=0.7\columnwidth]{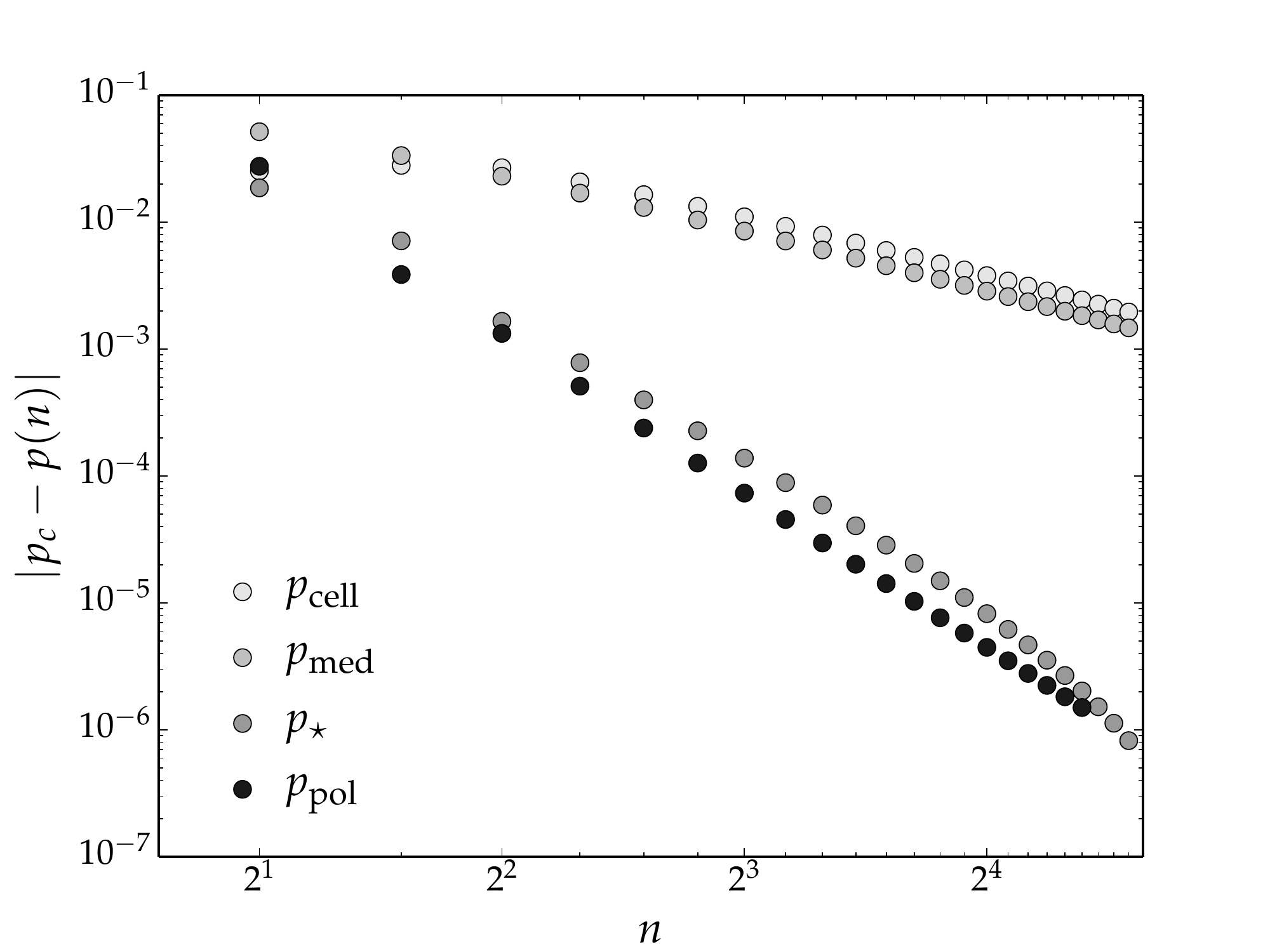}
  \caption{Finite size estimators for $p_c$ and their convergence.}
  \label{fig:pcestimators}
\end{figure}

Both estimators are supposed to converge with the same leading exponent,
\begin{equation}
  \label{eq:leading-convergence}
  \pmed(n)-p_c \simeq c_{\mathrm{med}}\, n^{-1-1/\nu} \qquad
  \pcell(n)-p_c \simeq c_{\mathrm{cell}} \, n^{-1-1/\nu}
\end{equation}
with $\nu=\frac{4}{3}$. This scaling can clearly be seen in the data (Figure~\ref{fig:pcestimators}).

Any convex combination of finite size estimators is another finite
size estimator. We can choose $\lambda$ in
$\lambda \pmed(n) + (1-\lambda) \pcell(n)$ such that the leading order
terms of $\pmed(n)$ and $\pcell(n)$ cancel:
\begin{equation}
  \label{eq:lambda-1}
  \lambda =\left(1-\frac{c_{\mathrm{med}}}{c_{\mathrm{cell}}}\right)^{-1}\,.
\end{equation}
The precise values of $c_{\mathrm{med}}$ and $c_{\mathrm{cell}}$ are not
known, but their ratio follows from scaling arguments \cite{ziff:newman:02}:
\begin{equation}
  \label{eq:c-ratio}
  \frac{c_{\mathrm{med}}}{c_{\mathrm{cell}}} = -\frac{1}{\nu} = -\frac{3}{4}\,.
\end{equation}
Hence we expect that the estimator
\begin{equation}
  \label{eq:pstar_def}
  \pstar(n) \equiv \frac{4}{7}\,\pmed(n) + \frac{3}{7}\,\pcell(n)
\end{equation}
converges faster than both $\pmed(n)$ and $\pcell(n)$. This is in fact
the case, as can be seen in Figure~\ref{fig:pcestimators}.

Figure~\ref{fig:pcestimators} also depicts the estimator $\ppol(n)$,
which is the root of a graph
polynomial defined for percolation on the $n\times n$ torus \cite{jacobsen:scullard:13,jacobsen:14,mertens:ziff:16}.
Empirically, it converges like $\sim n^{-4}$ in leading order.
Jacobsen \cite{jacobsen:15} computed this estimator for $n\leq 21$ and
extrapolated the results to obtain
\begin{equation}
  \label{eq:pc_jacobsen}
  p_c= \numprint{0.59274605079210}(2) \,.
\end{equation}
This is the most precise value up to today. We have used it as
reference $p_c$ in Figure~\ref{fig:pcestimators} and we will continue to
use it as reference in the rest of the paper.

\subsection{Exponents}

Extrapolation methods are commonly based on the assumption of a power law scaling
\begin{equation}
  \label{eq:ansatz_estimator}
  p(n) = p_c + \sum_{k=1} A_k n^{\Delta_k}
\end{equation}
with exponents $0 >\Delta_1 > \Delta_2 > \ldots$.
These exponents can be determined by the following procedure.

\begin{figure}
  \centering
  \includegraphics[width=\columnwidth]{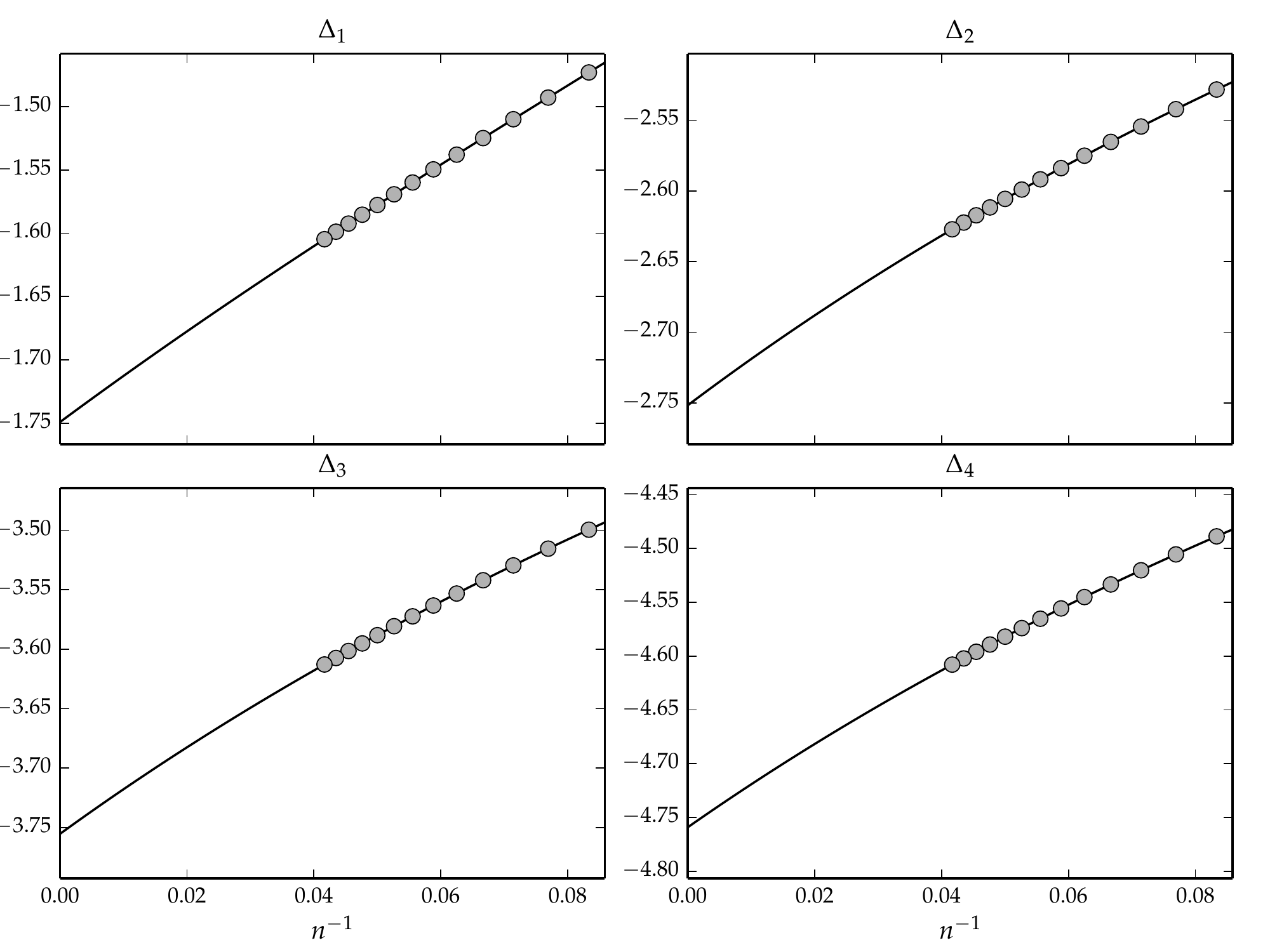}
  \caption{Exponent sequences $\Delta_k(n)$ of $\pmed(n)$ and
    4th-order polynomial fits.}
  \label{fig:exponents}
\end{figure}

We start by considering the truncated form
$p(n)=p_c+A_1n^{\Delta_1}$. From each triple of consecutive data points
$p(n)$, $p(n-1)$ and $p(n-2)$ we can compute three unknows
$p_c$, $A_1$ and $\Delta_1$. This gets us sequences $A_1(n)$ and
$\Delta_1(n)$ that we can use to compute
$\Delta_1=\lim_{n\to\infty}\Delta_1(n)$ and $A_1 \equiv \lim_{n\to\infty}A_1(n)$.
Figure~\ref{fig:exponents} shows $\Delta_1(n)$ for $p(n)=\pmed(n)$
plotted versus $n^{-1}$ along with a 4th-order
polynomial fit. The result $\Delta_1 = -\frac{7}{4}$
is of course expected. It is stable under 
variations of the order of the fit and the number of low
order data points excluded from the fit.

We then subtract the sequence $A_1 n^{\Delta_1}$ from $p(n)$ to
eliminate the leading scaling term $n^{\Delta_1}$. Considering again a
truncated form $p(n)-A_1n^{\Delta_1} = A_2n^{\Delta_2}$, we can
get sequences $A_2(n)$ and $\Delta_2(n)$ from three consecutive terms
of this sequence. And again we extrapolate these sequences by fitting
a low order polynomial in $n^{-1}$, thereby obtaining $\Delta_2$ and
$A_2$.

Obviously, this can be iterated to determine the exponents
$\Delta_k$ one by one, allthough the stability of the procedure deteriorates to
some extent with each iteration. Figure~\ref{fig:exponents} shows the
extrapolations for $\Delta_2$, $\Delta_3$ and $\Delta_4$ for $\pmed$
with results
\begin{equation}
   \label{eq:Deltas}
   \Delta_2 = -2.7516 \approx -\frac{11}{4} \quad
   \Delta_3 = -3.7553 \approx -\frac{15}{4} \quad
   \Delta_4 = -4.7590 \approx -\frac{19}{4}
\end{equation}
The results for $\pcell$ are similar. These results provide compelling
evidence that
\begin{equation}
  \label{eq:exponents_pmed_pcell}
  \Delta_k = -\frac{4k+3}{4}\qquad(\pmed, \pcell)\,.
\end{equation}
These are the exponents that we will use in our
extrapolations.

Applying the same procedure,  Jacobsen \cite{jacobsen:15} concluded that the
scaling exponents for $\ppol$ are
\begin{equation}
  \label{eq:exponents_ppol}
  \Delta_k = -2k-2 \qquad (\ppol)\,.
\end{equation}

For $\pstar(n)$, we could not reliably compute exponents $\Delta_k$.
It can already be sensed from the curvature in the logarithmic plot
(Figure~\ref{fig:pcestimators}), that $\pstar(n)$ is not determined by
a set of well separated exponents. This is the reason why we will not
use $\pstar(n)$ for extrapolation in this contribution, despite its
fast convergence.

\subsection{Extrapolation}

A study of extrapolation methods in statistical physics
\cite{henkel:schuetz:88} demonstrated that Bulirsch-Stoer (BST)
extrapolation \cite{bulirsch:stoer:64} is a reliable, fast converging
method.  It is based on rational interpolation. For a given sequence
of exponents $0 < w_1 < w_2 < \ldots$ and $n$ data points
$p(1),\ldots,p(n)$, one can compute coefficients $a_0,\ldots,a_N$ and
$b_1,\ldots,b_M$ with $N+M+1=n$ such that the function
\begin{equation}
  \label{eq:def_rational_function}
  Q_{N,M}(h) = \frac{a_0+a_1h^{w_1}+a_2h^{w_2}+\cdots+a_Nh^{w_N}}{1+b_1h^{w_1}+b_2h^{w_2}+\cdots+b_Mh^{w_M}}
\end{equation}
interpolates the data: $p(k)=Q_{N,M}(1/k)$ for $k=1,\ldots,n$.
The value of $a_0$ yields the desired extrapolation $p(k\to\infty)$.

The best results are usually obtained with ``balanced''\  rationals
$N=\lceil (n-1)/2\rceil$ and $M=\lfloor(n-1)/2\rfloor$.
This is what we will use for our extrapolations.

The BST method discussed in \cite{henkel:schuetz:88} uses
exponents $w_k = k\omega$, where the parameter $\omega$ is chosen to
match the leading order of the finite size corrections. Here we choose
$w_k=-\Delta_k = \frac{4k+3}{4}$.

\begin{figure}
  \centering
  \includegraphics[width=0.7\columnwidth]{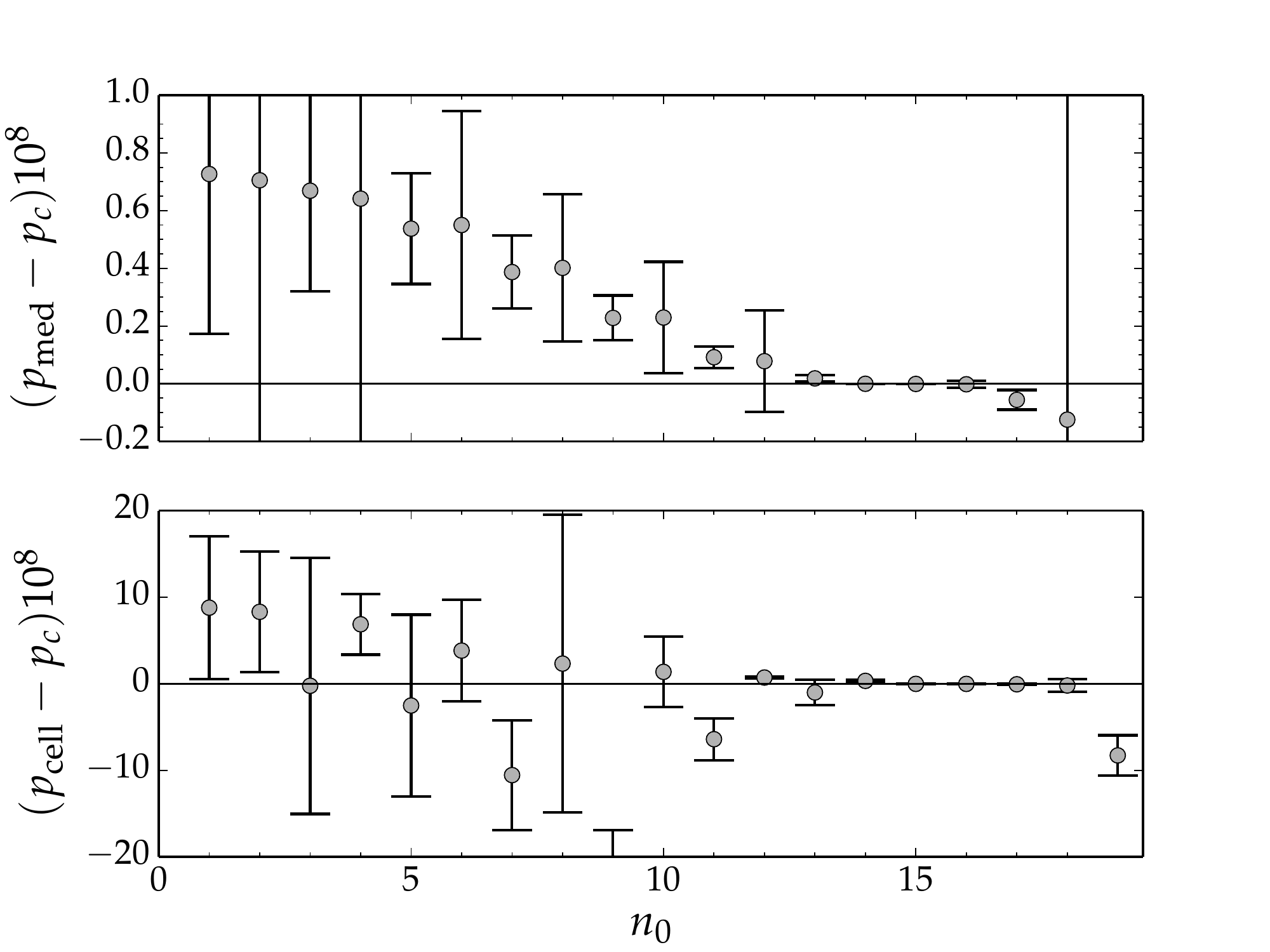}\\
  \caption{Rational extrapolation result vs. minimum system
    size $n_0$.}
  \label{fig:pmed_pcell_estimates}
\end{figure}

Having fixed the exponents and the degrees of nominator and
denominator, we still have a free parameter: the number of data points
to be included in the extrapolation. Since we are interested in the
asymptotics $n\to\infty$ we will of course use the data of the large systems,
but we may vary the size $n_0$ of the smallest system to be included.
We expect that the extrapolation gets more accurate with increasing
$n_0$, but only up to a certain point after which the quality
deteriorates because we have to few data points left. This intuition
is confirmed by the data (Figure~\ref{fig:pmed_pcell_estimates}).

The accuracy of an extropolation based on system sizes
$n_0,\ldots,\nmax$ is gauged by the difference between extrapolations of
data points for $n_0+1,\ldots,\nmax$ and $n_0,\ldots,\nmax-1$. These
are the errorbars shown in Figure~\ref{fig:pmed_pcell_estimates}.

\begin{table}
  \centering
  \def\arraystretch{1.5}
  \begin{tabular}{lll}
    $n_0$ & \pmed & \pcell \\\hline
    13 & \numprint{0.59274605098}(12)     & \numprint{0.592746041}(15)\\
    14 & \numprint{0.5927460507885}(22)$^\star$ & \numprint{0.5927460541}(16)\\
    15 & \numprint{0.592746050783}(6)$^\star$      & \numprint{0.592746050611}(7)$^\star$ \\
    16 & \numprint{0.59274605077}(12)     & \numprint{0.59274605059}(16)$^\star$ \\
    17 & \numprint{0.59274605023}(34)     & \numprint{0.5927460501}(5)
  \end{tabular}
  \caption{Extrapolations for different minimum system sizes
    $n_0$. Starred values are averaged to yield the final estimates \eqref{eq:mean_estimates} for
  $p_c$.}
  \label{tab:rational-extrapolations}
\end{table}

It seems reasonable to focus on the regime of $n_0$ where the results
form a plateau with small errorbars, and to take the mean of these
values as the final estimate of $p_c$. Using the values
for $n_0=14, 15$ ($\pmed$) and $n_0=15,16$ ($\pcell$)  (see
Table~\ref{tab:rational-extrapolations}) we get
\begin{subequations}
  \label{eq:mean_estimates}
\begin{align}
  \label{eq:mean_pmed}
  p_c &= \numprint{0.592746050786}(3) \qquad (\pmed)\,,\\
  \label{eq:mean_pcell}
  p_c &= \numprint{0.59274605060}(8)  \qquad (\pcell)\,.
\end{align}
\end{subequations}
The first value deviates from the reference value
\eqref{eq:pc_jacobsen} by 2 errorbars, the second by 2.5 errorbars.

\begin{figure}
  \centering
  \includegraphics[width=0.7\columnwidth]{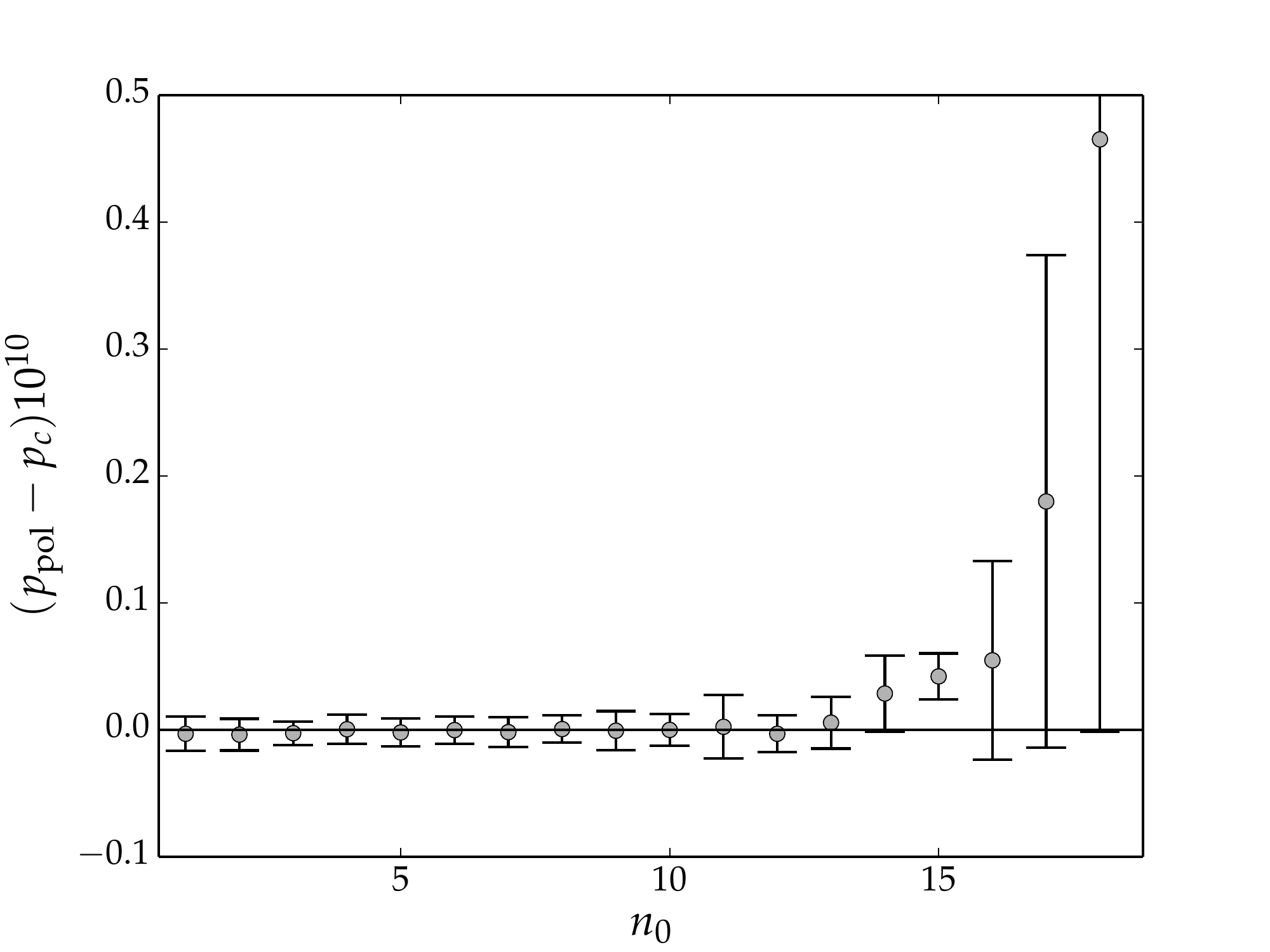}\\
  \caption{Rational extrapolation $\ppol$ vs. minimum system
    size $n_0$.}
  \label{fig:ppol_estimates}
\end{figure}

The reference value $p_c$ \ref{eq:pc_jacobsen} was obtained with
different extrapolation approach. To see how consistent our method is,
we applied it also to the estimator $\ppol(n)$ from
Jacobsen \cite[Table.~2]{jacobsen:15}. Here we used the
exponents \eqref{eq:exponents_ppol} in the BST extrapolation.

Figure~\ref{fig:ppol_estimates} shows, that the errorbars are smaller
than the errorbars on $\pmed$ and $\pcell$ (note the scaling of the
$y$-axis). The estimator $\ppol$ is also more robust under variation
of $n_0$. Taking the average of the estimates for $n_0=1,\ldots,10$
gives
\begin{equation}
  \label{eq:equation}
  p_c = \numprint{0.5927460507920}(4) \quad(\ppol)\,.
\end{equation}
This value agrees with Jacobsen's value \eqref{eq:pc_jacobsen} within
the errorbars.

We believe that the superiority of the estimator $\ppol$ is due to the
better separation of scales: $\Delta_{k+1}-\Delta_{k} = -1$ for $\pmed,
\pcell$ versus $\Delta_{k+1}-\Delta_{k} = -2$ for $\ppol$.

%% file: conclusions.tex
\section{Conclusions}
\label{sec:conclusions}

We have introduced a transfer matrix algorithm to compute the
generating function $F_{nm}(z)$ for the number of spanning configurations in an
$n\times m$ square lattice. Adapting this algorithm to other $2d$
lattices and other boundary conditions (cylindrical, helical) is
straightforward.

The exponential number of signatures imposes a hard limit on system
sizes.  With the current algorithm, computing $F_{25,25}(z)$ would
require a computer with a 4 TB memory and a couple of months CPU time.
Evaluating $R_{25,25}(p)$ for a single real argument $p$ requires 512 GB
of memory und would take about one month.

We have also seen another example where the exact solution of small systems
is sufficient to compute estimates for the critical density $p_c$
with an accuracy that no Monte-Carlo method could ever
achieve. Or, as Bob Ziff put it, ``Monte-Carlo is dead''
\cite{ziff:talk:21} when it comes to computing $p_c$ for $2d$ systems.

This approach can be advanced. Not so much by solving larger systems,
but by a better understanding of finite size estimators and their
scaling.

\section*{Acknowledgements}

This paper benefitted a lot from discussions with Bob Ziff. Like
almost all of my previous papers on percolation! Whenever I had a question
or a crazy idea on percolation, I could ask Bob. And I always got a
quick and satisfying answer.

It is a great pleasure to thank Bob for his support and his
friendship.

%% file: nsig.tex
\section{The Number of Signatures}
\label{sec:nsig}

To compute the number $S_n$ of signatures in fully completed
rows, we use the connection of signatures and balanced strings of parentheses.

The number of balanced strings of parentheses length $2a$ is given by the Catalan number $C_a$ \cite[Theorem
7.3]{roman:catalan},
\begin{equation}
  \label{eq:def-catalan}
  C_a = \frac{1}{a+1}\,{2a \choose a} = \frac{(2a)!}{(a+1)! a!} =
  \prod_{k=2}^a \frac{a+k}{k}\,.
\end{equation}
A completely balanced string of parantheses contains the same number
$a$ of opening and closing parentheses. We will see that it is useful
to consider also strings of parentheses that consist of $a$ opening
parentheses and $b \leq a$ closing parentheses such that reading from
left to right, every closing parenthesis matches an opening
parenthesis. The number of these partially balanced strings of
parentheses reads \cite{bailey:96}
\begin{equation}
  \label{eq:def-catalan-triangle}
  C_{a,b} = \frac{a-b+1}{a+1}{a+b \choose b}\qquad (a \geq b \geq 0)\,.
\end{equation}
The numbers $C_{a,b}$ are known as Catalan's triangle, and
the Catalan numbers are the diagonal $C_a =
C_{a,a,}$ of this triangle.

\subsection{From Catalan to Motzkin numbers}

In a signature, the parentheses are not independent.
Occupied sites that are adjacent in a row belong to the same
cluster.  We deal with this local connectivity by considering
runs of adjacent occupied sites as single entities that we may call
h(orizontal)-clusters.  A row of width $n$ contains at most
$\lceil n/2\rceil$ h-clusters. Different h-clusters may be connected
through previous rows, and again we have a unique left-right order and
the topoligical constraints that allow us to identify h-clusters with
strings of balanced parentheses expression, using the set of symbols
((, )), )( and (). Any legal signature with $r$ h-clusters
corresponds to a balanced string of $2r$ parentheses and vice versa.

We write the occupation pattern of a row as $n$-bit word
 $s\in\{0,1\}^n$, where $0$ represents an empty site and $1$
 represents an occupied site. Let $r(s)$ denote the number of
 h-clusters, i.e. runs of consecutive $1$s, in $s$. Then the
 number of signatures is
\begin{displaymath}
  \sum_{s\in\{0,1\}^n} C_{r(s)}\,.
\end{displaymath}
This sum can be simplified by computing the number of
$n$-bit words with exactly $r$ h-clusters. For that we  
add a leading zero to the word. This ensures that every h-cluster has a
zero to its left. The position of this zero and the position of the
rightmost site of the h-cluster specify the h-cluster uniquely. Hence the
number of $n$-bit words with $r$ h-clusters 
is given by ${{n+1} \choose {2r}}$, providing us with
\begin{equation}
  \label{eq:free-signatures}
  \sum_{r=0}^{\left\lfloor\frac{n+1}{2}\right\rfloor}
  {{n+1}\choose {2r}}\, C_r \equiv M_{n+1}\,,
\end{equation}
where $M_n$ is known as the $n$th Motzkin number
\cite{motzkin:48,oste:vanderjeugt:15}.

\subsection{The spanning constraint}

\begin{figure}
  \centering
  \includegraphics[width=0.7\columnwidth]{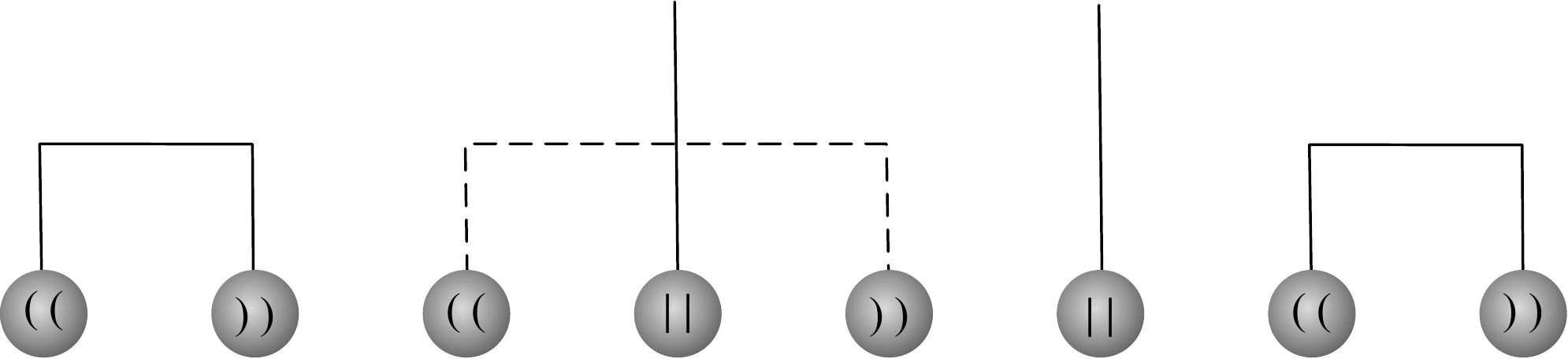}
  \caption{The dashed connection is forbidden: h-clusters can not connect
    across h-clusters that are connected to the top row.}
  \label{fig:signature-forbidden}
\end{figure}

Up to this point we have ignored that our signatures have an additional symbol
$\vert\vert$ for occupied sites connected to the top row.  In a
signature with $r$ h-clusters, any number $w \leq r$ of h-clusters can be connected
to the top row:
\begin{equation}
  \label{eq:spanning-signatures-ub}
  S_n \leq
  \sum_{r=1}^{\left\lceil\frac{n}{2}\right\rceil}
  {{n+1}\choose {2r}}\, \sum_{w=1}^r {r \choose w} \, C_{r-w}\,.
\end{equation}
This is only an upper bound because it does not take into
account that h-clusters not connected to the top row can not be
connected to each other across a top row h-cluster. In the example
shown in Figure~\ref{fig:signature-forbidden}, the number of signatures
compatible with the top row h-cluster is not $C_{6}=132$ but 
$C_{3} C_{1} C_{2} = 10$. 
 
We need to consider the compositions that $w$ top row h-clusters can
impose upon the $r-w$ other h-clusters.  
The composition of an integer $n$ into $k$ parts is 
defined as an ordered arrangement of $k$ non-negative integers which sum
to $n$ \cite{heubach:mansour:09}. The compositions of $4$ into $2$ parts are
\begin{displaymath}
  (0,4),  (1,3), (2,2), (3,1), (4,0)\,.
\end{displaymath}
A number of $w$ top row h-clusters can divide the $r-w$ other h-clusters into
$w+1$ parts at most. This gets us
\begin{displaymath}
  S_n =
  \sum_{r=1}^{\left\lceil\frac{n}{2}\right\rceil}
  {{n+1}\choose {2r}}\, \sum_{w=1}^r 
  \sum_{\substack{i_k \geq 0 \\i_1+\ldots+i_{w+1}=r-w}} C_{i_1}\cdots C_{i_{w+1}}\,.
\end{displaymath}
Using Catalan's convolution formula \cite{regev:12} for $1 \leq k \leq n$,
\begin{equation}
  \label{eq:catalan-convolution}
  \sum_{i_1+\ldots +i_k=n} C_{i_1-1} \cdots
  C_{i_k-1}=\frac{k}{2n-k}{{2n-k}\choose n} \qquad (1\leq k \leq n)\,,
\end{equation}
we can write this as
\begin{equation}
  \label{eq:spanning-signatures-1}
  S_n =
  \sum_{r=1}^{\left\lceil\frac{n}{2}\right\rceil}
  {{n+1}\choose {2r}}\, \sum_{w=1}^r 
  \frac{w+1}{2r-w+1}{{2r-w+1}\choose {r+1}}\,.
\end{equation}
The sum over $w$ can be done using binomial coefficient
identities \cite[Table 174]{graham:knuth:patashnik:94}. The result reads
\begin{align}
  S_n &=
  \sum_{r=1}^{\left\lceil\frac{n}{2}\right\rceil}
  {{n+1}\choose {2r}}\, \frac{3}{r+1}{{2r} \choose {r-1}}\nonumber\\
  &= \sum_{r=1}^{\left\lceil\frac{n}{2}\right\rceil}
  {{n+1}\choose {2r}}\, C_{r+1,r-1}\,,  \label{eq:spanning-signatures-2}
\end{align}
where \eqref{eq:spanning-signatures-2} follows from \eqref{eq:def-catalan-triangle}
with $a=r+1$ and $b=r-1$.
This result can be categorized by considering the generalization of
\eqref{eq:free-signatures}. Analog to Catalan's triangle, Motzkin's triangle
\cite{donaghey:shapiro:77} is defined as
\begin{equation}
  \label{eq:motzkin-catalan-triangle}
  M_{n,k} = \sum_{r=0}^{\lfloor \frac{n-k}{2}\rfloor} {n \choose {2r+k}}\,C_{r+k,r}.
\end{equation}
Hence $S_n=M_{n+1,2}$, the second column in Motzkin's triangle. This
sequence is number \seqnum{A005322} in the on-line
encyclopedia of integer sequences.

\subsection{Asymptotics}

Allthough Motzkin's triangle is a well known object in
combinatorics, we could not find an explicit formula for the
asymptotic scaling of $M_{n,2}$. Hence we compute it here, following
the procedure outlined in \cite{flajolet:sedgewick:book}.

Our starting point is the the generating function for $S_n=M_{n+1,2}$ \cite{donaghey:shapiro:77}
\begin{equation}
  \label{eq:Motzkin_column_generating}
  F(x) = \sum_{n>0} S_n \,x^{n} = \frac{\left((1-x-\sqrt{(1+x)(1-3x)}\right)^3}{8x^5}\,. 
\end{equation}
The pole that determines the asymptotics of the coefficients is
$x=1/3$. Expanding $F$ in terms of $\sqrt{1-3x}$ gives
\begin{displaymath}
  F(x) = 9-27\sqrt{3}\, (1-3x)^{\frac{1}{2}} + \frac{279}{2} (1-3x) -
  \frac{1485\sqrt{3}}{8} (1-3x)^{\frac{3}{2}} + \cdots\,.
\end{displaymath}
The coefficients of the binonmial series that appear in this expansion
are
\begin{equation}
  \label{eq:bionomial}
  \begin{aligned}
  \left[x^n\right](1-3x)^{\alpha} &= (-3)^n{\alpha \choose n} =
  3^n {{n-\alpha -1}\choose n}\\
  &=
  3^n \frac{(-\alpha)(1-\alpha)\cdots(n-1-\alpha)}{n!}\,.
  \end{aligned}
\end{equation}
For $\alpha = 0,1,2,3,\ldots$, the binomial series is finite and
does not contribute to the asymptotic scaling. For 
$\alpha\neq 0,1,2,3,\ldots$, the binomial coefficient can be written as
\begin{equation}
  \label{eq:bionomial-gamma}
  {{n-\alpha -1}\choose n} = \frac{\Gamma(n-\alpha)}{\Gamma(-\alpha)\Gamma(n+1)}\,,
\end{equation}
and we can use the asymptotic series for the $\Gamma$ function
\begin{equation}
  \label{eq:gamma-asympt}
   \begin{aligned}
  \frac{\Gamma(n-\alpha)}{\Gamma(n+1)} \sim 
  n^{-\alpha}
   \left(\frac{1}{n}+\frac{\alpha(\alpha+1)}{2n^2}
      +\frac{\alpha(\alpha+1)(\alpha+2)(3\alpha+1)}{24n^3}
    \right.
    \\
  \left.+\frac{\alpha^2(\alpha+1)^2(\alpha+2)(\alpha+3)}{48n^4} +O\left(\frac{1}{n^5}\right)   \right)\,.
\end{aligned}
\end{equation}
to obtain
\begin{equation}
  \label{eq:aymptotics_generating_function}
  \left[x^n\right]F(x)\sim\frac{\sqrt{3}\,3^{n+3}}{\sqrt{4\pi
      n^3}}\left(1-\frac{159}{16 n} + \frac{36505}{512 n^2} +O(\frac{1}{n^3})\right)
\end{equation}
